\documentstyle[aps,epsf,preprint]{revtex}

\title{Teleportation, Entanglement and Thermodynamics in 
the Quantum World}
\author{Martin B. Plenio and Vlatko Vedral$^{*}$\\
{\protect\small\em $^{1}$ Blackett Laboratory, Imperial College, 
Prince Consort Road, London SW7 2BZ, U.K.}}

\date{\today}

\begin{document}
\maketitle

\begin{abstract}
Quantum mechanics has many counter-intuitive consequences which 
contradict our intuition which is based on classical physics. 
Here we discuss a special aspect of quantum mechanics, namely 
the possibility of entanglement between two or more particles. 
We will establish the basic properties of entanglement using 
quantum state teleportation. These principles will then allow us 
to formulate quantitative measures of entanglement. 
Finally we will show
that the same general principles can also 
be used to prove seemingly 
difficult questions regarding entanglement dynamics
very easily. This will be 
used to motivate the hope that we can 
construct a thermodynamics of entanglement.        
\end{abstract}
\vfill
* present address: Clarendon Laboratory, University of Oxford, Parks Road,
Oxford OX1 3PU, UK
\newpage

\section{Introduction}
Quantum mechanics is a non-classical theory and therefore exhibits many 
effects that are counter-intuitive. This is because in our everyday 
life we experience a classical (macroscopic) world with respect to which we
define ``common sense". One principle that lies at the heart of quantum mechanics
is the superposition principle. In itself it might still be understood within
classical physics, as it crops up e.g. in classical electrodynamics. However, 
unlike in classical theory the superposition principle in quantum mechanics 
also gives rise to a property called entanglement between quantum mechanical
systems. This is due to the Hilbert space structure of the quantum mechanical
state space. In classical mechanics particles can be correlated
over long distances simply because one observer can prepare a system in a
particular state and then instruct a different observer to prepare the same state. 
However, all the correlations generated in this way can be understood perfectly 
well using classical probability distributions and classical intuition. The 
situation changes dramatically when we consider correlated systems in quantum 
mechanics. In quantum mechanics we can prepare two particles in such a way that 
the correlations between them cannot be explained classically. Such states
are called entangled states. It was the great 
achievement of Bell to recognize this fact and to cast it
into a mathematical form that, in principle, allows the test of quantum mechanics
against local realistic theories \cite{Bell65,Bell66,Bell87,Clauser78}. Such 
tests have been performed, and the quantum mechanical predictions have been 
confirmed \cite{Aspect82} although it should be noted that an experiment that
has no loopholes (these are insufficiencies in the experiment that allow the
simple construction of a local hidden variable theory) has not yet been performed
\cite{Santos}. With the formulation of the Bell inequalities and the experimental 
demonstration of their violation, it seemed that the question of the non-locality 
quantum mechanics had been settled once and for all. However, in recent 
years it turned out this conclusion was premature. While indeed the 
entanglement of pure states can be viewed as well understood, the 
entanglement of mixed states still has many properties that are 
mysterious, and in fact new problems (some of which we describe here) 
keep appearing. The reason for the problem with mixed states lies in 
the fact that the quantum content of the correlations is hidden behind 
classical correlations in a mixed state.
One might expect that it would be impossible to recover the quantum content of
the correlations but this conclusion would be wrong. Special methods have been 
developed that allow us to 'distill' out the quantum content of the correlations
in a mixed quantum state \cite{Bennett96a,Bennett96b,Deutsch96,Gisin96,Horodecki97}. 
In fact, these methods showed that a mixed state which does not violate Bell inequalities 
can nevertheless reveal quantum mechanical correlations, as one can distill from
it pure maximally entangled states that violate Bell inequalities. Therefore,
Bell inequalities are not the last word in the theory of quantum entanglement. This has opened
up a lot of interesting fundamental questions about the nature of entanglement
and we will discuss some of them here. We will study the problem of 
how to quantify entanglement \cite{Bennett96c,Vedral97a,Vedral97c,VP98a}, the 
fundamental laws that govern entanglement transformation and the connection of
these laws to thermodynamics. 

On the other hand, the new interest in quantum entanglement has also 
been triggered by the discovery that it allows us 
to transfer (teleport) an unknown quantum state 
of a two-level system from one particle to another distant particle without 
actually sending the particle itself \cite{Bennett93}. 
As the particle itself is not sent, this represents a method of
secure transfer of information from sender to receiver 
(commonly called Alice and Bob), and eavesdropping is impossible. The key 
ingredient in teleportation is that 
Alice and Bob share a publicly known maximally entangled state 
between them. To generate such a state in practise one 
has to employ methods of quantum state distillation as mentioned above
which we review in Section 3. The protocol of quantum teleportation
has been recently implemented experimentally using single photons in laboratories
in Innsbruck \cite{Zeilinger98} and Rome \cite{deMartini98}, which
only adds to the enormous excitement that the field of
quantum information is currently generating.

But perhaps the most spectacular application of 
entanglement is the quantum computer, which could allow, once 
realized, an exponential increase of computational speed
for certain problems such as for example the factorization of 
large numbers into primes, for further explanations see the 
reviews \cite{VP98b,Ekert96,Barenco96}. Again at the heart of 
the idea of a quantum computer lies the principle of entanglement. 
This offers the possibility of massive parallelism in quantum systems 
as in quantum mechanics $n$ quantum systems can 
represent $2^n$ numbers simultaneously \cite{Ekert96,Jozsa97,VP98b}. The 
disruptive influence of the environment makes the realization 
of quantum computing
extremely difficult \cite{Plenio96,Plenio97} and many ideas have been developed to 
combat the noise in 
a quantum computer, incidentally again using entanglement  
\cite{Calderbank96,Chiara96,Shor95,Steane96}. Many other applications
of entanglement are now being developed and investigated, 
e.g. in frequency standards
\cite{Huelga97}, distributed quantum computation \cite{Grover97,Huelga98},
multiparticle entanglement swapping \cite{Bose98} and multiparticle
entanglement purification \cite{Murao98}.

In this article we wish to explain the basic ideas and problems 
behind quantum 
entanglement, address some fundamental questions and present 
some of its 
consequences, such as teleportation and its use in (quantum) communication.
Our approach is somewhat unconventional. Entanglement is
usually introduced through quantum states which violate the classical
locality requirement (i.e. violate Bell's inequalities) as we have
done above. 
Here we abandon this approach altogether and 
show that there is much more 
to entanglement than the issue of locality. 
In fact, concentrating on other
aspects of entanglement helps us to  
view the nature of quantum 
mechanics from a different angle. We hope that
the reader will, after studying this article, share our 
enthusiasm for the problems of the new and rapidly expanding
field of quantum information theory, at the heart of which lies
the phenomenon of quantum correlations and entanglement.

\section{Quantum Teleportation}

We first present an example that crucially depends on the 
existence of quantum mechanical correlations, i.e. entanglement. 
The procedure 
we will analyse is called quantum teleportation and can 
be understood as follows. The naive idea of teleportation involves a
protocol whereby an object positioned at a place $A$ and
time $t$ first ``dematerializes" and then reappears at a distant
place $B$ at some later time $t+T$. Quantum teleportation 
implies that we wish to apply this procedure to a quantum 
object. However, a genuine quantum teleportation differs from 
this idea, because we
are not teleporting the whole object but just its state from particle
$A$ to particle $B$. As
quantum particles are indistinguishable anyway, this amounts to `real'
teleportation. One way of performing teleportation (and certainly the 
way portrayed in various science fiction movies, e.g. The Fly) is first to learn 
all the properties of that object (thereby possibly destroying it). 
We then send this information as a classical string of data to $B$
where another object with the same properties is re-created.
One problem with this picture is that, if we have a single quantum
system in an unknown state, we cannot determine its state
completely because of the uncertainty principle. More precisely,
we need an infinite ensemble of identically prepared quantum systems
to be able completely to determine its quantum state. So it
would seem that the laws of quantum mechanics prohibit
teleportation of single quantum systems. 
However, the very feature of quantum mechanics
that leads to the uncertainty principle (the superposition principle) 
also allows the existence
of entangled states. These entangled states will provide a 
form of quantum channel to conduct a teleportation protocol. 
It will turn out that there is no need to learn the state
of the system in order to teleport it. On the other hand, 
there is a need to send some classical information from $A$ to $B$, but 
part of the information also travels down an entangled channel.
This then provides a way of distinguishing quantum and 
classical correlations, which we said was at the heart 
of quantifying entanglement.
After the teleportation is completed, the original state of the 
particle at $A$ is destroyed
(although the particle itself remains intact) and so is the 
entanglement in the quantum channel. These two
features are direct consequences of fundamental laws that are
central for understanding entanglement  
as we explain in a more detail in the next subsection.    

\subsection{A basic description of teleportation}

Let us begin by describing quantum teleportation in the form originally
proposed by Bennett et al \cite{Bennett93}. 
Suppose that Alice and Bob, who are distant from each other,
wish to implement a teleportation procedure. Initially they need
to share a maximally entangled pair of quantum mechanical 
two level systems. A two level system 
in quantum mechanics is also called a quantum bit, or qubit 
\cite{Schumacher95}, in
direct analogy with the classical bit of information (which is
just two distinguishable states of some system). Unlike the
classical bit, a qubit can be in a superposition of
its basis states, like $|\Psi\rangle = a|0\rangle + b|1\rangle$. 
This means that if Alice and Bob both have one qubit each then  
the joint state may for example be
\begin{equation}
|\Psi_{AB}\rangle = (|0_A\rangle |0_B\rangle + |1_A\rangle |1_B\rangle)/\sqrt{2}\; ,
\end{equation}
where the first ket (with subscript A) belongs to Alice and second 
(with subscript B) to Bob. This
state is entangled meaning, that it cannot be written as a 
product of the individual states (like e.g. $|00\rangle$). 
Note that this state is different from a statistical mixture 
$(|00\rangle\langle 00| + |11\rangle\langle 11|)/2$ which is the
most correlated state allowed by classical physics. 

Now
suppose that Alice receives a qubit in a state which is unknown
to her (let us label it $|\Phi\rangle = a|0\rangle + b|1\rangle$) and
she has to teleport it to Bob. 
The state has to be unknown to her because otherwise 
she can just phone Bob up and tell him all the details of the
state, and he can then recreate it on a particle that he 
possesses. If Alice does not know the state, then she cannot
measure it to obtain all the necessary information to specify it. Therefore
she has to resort to using the state $|\Psi_{AB}\rangle$ that
she shares with Bob. To see what she has to do, we write out
the total state of all three qubits
\begin{equation}
|\Phi_{AB}\rangle := |\Phi\rangle |\Psi_{AB}\rangle = (a|0\rangle + b|1\rangle)(|00\rangle + |11\rangle)/\sqrt{2} \;\; .
\end{equation}
However, the above state can be written in the following 
convenient way (here we are only rewriting the above expression in a different basis,
and there is no physical process taking place in between)
\begin{eqnarray}
|\Phi_{AB}\rangle & = & (a|000\rangle + a|011\rangle + b|100\rangle + b|111\rangle)/\sqrt{2} 
\nonumber \\
& = & \frac{1}{2}\left[ |\Phi^+\rangle (a|0\rangle + b|1\rangle) + |\Phi^-\rangle 
(a|0\rangle - b|1\rangle) + 
|\Psi^+\rangle (a|1\rangle + b|0\rangle) + 
|\Psi^-\rangle (a|1\rangle - b|0\rangle) \right] \; , 
\end{eqnarray}
where 
\begin{eqnarray}
|\Phi^+\rangle &=& (|00\rangle + |11\rangle)/\sqrt{2} \label{Bell1}\\
|\Phi^-\rangle &=& (|00\rangle - |11\rangle)/\sqrt{2} \\ 
|\Psi^+\rangle &=& (|01\rangle + |10\rangle)/\sqrt{2} \\ 
|\Psi^-\rangle &=& (|01\rangle - |10\rangle)/\sqrt{2}  \;  
\label{Bell}
\end{eqnarray}
form an ortho-normal basis of Alice's two qubits (remember that
the first two qubits belong to Alice and the last qubit belongs
to Bob). The above basis is frequently called the Bell basis. 
This is a
very useful way of writing the state of Alice's two qubits and
Bob's single qubit because it displays a high degree of
correlations between Alice's and Bob's parts: to every state 
of Alice's two qubits (i.e. $|\Phi^+\rangle, |\Phi^-\rangle,
|\Psi^+\rangle, |\Psi^-\rangle$) corresponds a state of
Bob's qubit. In addition the state of Bob's qubit in all
four cases looks very much like the original qubit that 
Alice has to teleport to Bob.  It is now straightforward to 
see how to proceed with the teleportation protocol \cite{Bennett93}:
\begin{enumerate}
\item Upon receiving the unknown qubit in state $|\Phi\rangle$ Alice performs
projective measurements on her two qubits in the Bell
basis. This means that she will obtain one of the 
four Bell states randomly, and with equal probability.

\item Suppose Alice obtains the state $|\Psi^+\rangle$. Then
the state of all three qubits (Alice $+$ Bob) collapses
to the following state
\begin{equation}
|\Psi^+\rangle (a|1\rangle + b|0\rangle)\; .
\end{equation} 
(the last qubit belongs to Bob as usual).
Alice now has to communicate the result of her measurement to Bob
(over the phone, for example). The point of this communication is 
to inform Bob how the state of his qubit now differs from the state 
of the qubit Alice was holding previously. 

\item Now Bob knows exactly what to do in order to 
complete the teleportation. He has to apply a unitary 
transformation on his qubit which simulates a logical 
NOT operation: $|0\rangle \rightarrow |1\rangle$ and
$|1\rangle \rightarrow |0\rangle$. He thereby transforms
the state of his qubit into the state $a|0\rangle + b|1\rangle$,
which is precisely the state that Alice had to teleport to him
initially. This completes the protocol. It is easy to see 
that if Alice obtained some other Bell state then Bob
would have to apply some other simple operation to complete
teleportation. We leave it to the reader to work out the other
two operations (note that if Alice obtained $|\Phi^+\rangle$ he would
not have to do anything). If $|0\rangle$ and $|1\rangle$ are 
written in their vector form then the operations that 
Bob has to perform can be represented by the Pauli spin matrices,
as depicted in Fig. \ref{teleport}. 

\end{enumerate}
An important fact to observe in the above protocol 
is that all the operations (Alice's measurements and Bob's
unitary transformations) are {\em local} in nature. This means 
that there is never any need to perform a (global) transformation
or measurement on all three qubits simultaneously, which is 
what allows us to call the above protocol a genuine teleportation. 
It is also important that the operations that Bob performs are 
independent of the state that Alice tries to teleport to Bob.
Note also that the classical communication from Alice to Bob
in step 2 above is crucial because otherwise the protocol
would be impossible to execute (there is a deeper
reason for this: if we could perform teleportation without 
classical communication then Alice could send messages to 
Bob faster than the speed of light, see e.g. \cite{Vedral97b}).   

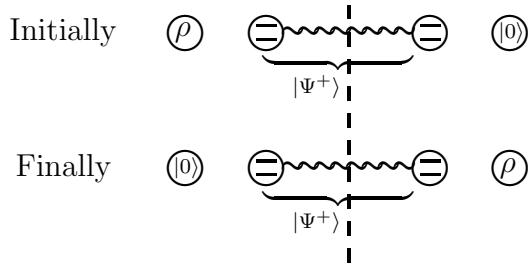
\begin{figure}[hbt]

\setlength{\unitlength}{0.9mm}
\begin{picture}(75,48)

\thicklines

\put(15,42){\makebox(0,0)[c]{Initially}}
\put(33,42){\circle{5}}
\put(45,42){\circle{5}}
\put(69.2,42){\circle{5}}
\put(81,42){\circle{5}}

\put(43.5,43){\line(1,0){3}}
\put(43.5,41){\line(1,0){3}}
\put(67.7,43){\line(1,0){3}}
\put(67.7,41){\line(1,0){3}}

\put(31.5,42){\makebox(0,0)[l]{$\small\rho$}}
\scriptsize
\put(79.3,42){\makebox(0,0)[l]{$ |0\rangle$}}
\normalsize

\put(44.5,37.5){\makebox(0,0)[l]{$\underbrace{\hspace{2.cm}}$ }}
\scriptsize
\put(53,34){\makebox(0,0)[c]{$|\Psi^{+}\rangle$ }}
\normalsize

\multiput(48.95,42.1)(2.35,0.){8}{\makebox(0,0)[c]{$\bf\sim$}}
\multiput(48.9,41.9)(2.35,0.){8}{\makebox(0,0)[c]{$\bf\sim$}}

\put(15,22){\makebox(0,0)[c]{Finally}}
\put(33,22){\circle{5}}
\put(45,22){\circle{5}}
\put(69.2,22){\circle{5}}
\put(81,22){\circle{5}}

\put(43.5,23){\line(1,0){3}}
\put(43.5,21){\line(1,0){3}}
\put(67.7,23){\line(1,0){3}}
\put(67.7,21){\line(1,0){3}}

\put(79.5,22){\makebox(0,0)[l]{$\small\rho$}}
\scriptsize
\put(31.3,22){\makebox(0,0)[l]{$ |0\rangle$}}
\normalsize

\put(44.5,17.5){\makebox(0,0)[l]{$\underbrace{\hspace{2.cm}}$ }}
\scriptsize
\put(53,14){\makebox(0,0)[c]{$|\Psi^{+}\rangle$ }}
\normalsize

\multiput(48.95,22.1)(2.35,0.){8}{\makebox(0,0)[c]{$\bf\sim$}}
\multiput(48.9,21.9)(2.35,0.){8}{\makebox(0,0)[c]{$\bf\sim$}}

\multiput(57.1,46)(0,-4){10}{\line(0,-1){2}}
\multiput(57.3,46)(0,-4){10}{\line(0,-1){2}}

\end{picture}
\caption{Again Alice is on the left of the dashed line and 
Bob on the right side. Assume that initially Alice and Bob are sharing two
particles in a maximally entangled state $|\psi\rangle$. Alice also holds a
particle in an unknown state $\rho$ while Bob holds a particle in the known 
state $|0\rangle$. The aim is that finally Alice and Bob have exchanged the
states of their particles {\bf and} that they are still sharing a pair 
of particles in the maximally entangled state $|\psi\rangle$. The question
whether this protocol is possible will be answered in Section V.}
\label{destroy1}
\end{figure}

Important to observe is also the fact that the initial state 
to be teleported is at the end destroyed, i.e it becomes
maximally mixed, of the form
$(|0\rangle\langle 0| + |1\rangle\langle 1|)/2$. This
has to happen since otherwise we would end up with two 
qubits in the same state at the end of teleportation (one
with Alice and the other one with Bob). So, effectively,
we would clone an unknown quantum state, which is impossible by the
laws of quantum mechanics (this is the no-cloning theorem of Wootters and Zurek 
\cite{Wootters82}).  
We also see that at the end of the protocol the quantum entanglement 
of $|\Psi_{AB}\rangle$ is completely destroyed. Does this have to be 
the case in general or might we save that state at the end
(by perhaps performing a different teleportation protocol)? Could we 
for example have a situation as depicted in Fig. \ref{destroy1}, where 
Alice teleports a quantum state from 
to Bob and afterwards the quantum channel is
still preserved. This
would be of great practical advantage, because we could
use a single entangled state over and over again to teleport an 
unlimited number of quantum states from Alice to Bob (this
question was first suggested to the authors by A. Ekert). 
Unfortunately the answer to the above question is NO:
the entanglement of the quantum channel has to be destroyed at the 
end of the protocol. The analytical proof of this seems to be extremely 
hard, because it appears that we have to check all the possible
purification protocols (infinitely many). However, the rest of 
this article introduces
new ideas and principles that will allow us to
explain more easily why this needs to be so. This explanation
will be presented at the end of this article.  
First, however, we need to understand why entanglement is necessary
for teleportation in the first place.

\subsection{Why is entanglement necessary?}

Quantum teleportation does not work if Alice and Bob share a
disentangled state. If we take that $|\Psi_{AB}\rangle = |00\rangle$ 
and run the same protocol as the above, then Bob's particle stays the 
same at the end of the protocol, i.e. there is no teleportation. 
In this case the total state of the three qubits would be
\begin{eqnarray}
|\Phi_1\rangle =  (a|0\rangle + b|1\rangle)|00\rangle \; .
\end{eqnarray}
We see that whatever we do (or, rather, whatever Alice does)
on the first two qubits and however
we transform them, the last qubit (Bob's qubit) will always
be in the state $|0\rangle$; it is thus completely
uncorrelated to Alice's two qubits and no teleportation is
possible. 

Thus one might be tempted to say that teleportation is unsuccessful 
because there are no correlations
between $A$ and $B$, i.e. $A$ and $B$ are statistically
independent from each other. 
So, let us therefore try a state of the form 
\begin{equation}
\rho_{AB} = 1/2 \left(|00\rangle\langle 00| + |11\rangle\langle 11| \right)\; .
\end{equation}
This state is a statistical mixture of the states $|00\rangle$ and $|11\rangle$,
both of which are disentangled. This is equivalent to Alice and
Bob sharing either $|00\rangle$ or $|11\rangle$, but being completely
uncertain about which state they have. This state is clearly correlated, because
if Alice has $0$ so does Bob, and if Alice has $1$ so does Bob.
However, since both the states are disentangled and neither one of them
achieves teleportation then their mixture cannot do it either. 
The interested reader can convince himself of this fact by
actually performing the necessary calculation, which is messy but
straightforward. It is important to stress that Alice is
in general allowed to perform any measurement on her qubits and 
Bob any state independent transformation on his qubit, but the teleportation would 
still not work with the above state \cite{Popescu94}. 
In fact, it follows that if $\{|\alpha^i_A\rangle\}$ is
a set of states belonging to Alice and $\{|\beta^j_B\rangle\}$ a set of states
belonging to Bob, then the most general state that cannot achieve
teleportation is of the form
\begin{equation}
\sigma_{AB} = \sum_{ij} p_{ij} |\alpha^i_A\rangle\langle\alpha^i_A| \otimes |\beta^j_B\rangle
\langle\beta^j_B| \; ,
\label{dis}
\end{equation}
where $p_{ij}$ are a set of probabilities such that $\sum_{ij} p_{ij}=1$.
This is therefore the most general disentangled state of two qubits. This 
state might have a certain amount of classical correlations as we 
have seen above, but any form of quantum correlations, i.e. entanglement, 
is completely absent \cite{Horodecki97}. So we can now summarize: both 
classical and quantum correlations are
{\em global} properties of two correlated systems, however, they can
be distinguished because classical correlations alone
cannot lead to teleportation. This establishes an important 
fact: entanglement plays a key role in the manipulation of
quantum information.

\subsection{The non-increase of entanglement under local operations}

The above discussion leads us to postulate one of the central laws
of quantum information processing. We now wish to encapsulate the
fact that if Alice and Bob share no entanglement they can
by no local means and classical communication achieve 
teleportation.  

\vspace*{1cm}
\noindent
\fbox{\parbox[b]{16cm}{ 
\begin{center} {\bf The fundamental law of quantum 
information processing.}
\end{center}
\vspace*{-0.3cm}
Alice and Bob cannot, with no matter how small a probability, by 
local operations and communicating  
classically turn a disentangled state $\sigma_{AB}$ into an 
entangled state.}}

\vspace*{1cm}
\noindent
The gist of the proof relies on reductio ad absurdum. Suppose they could turn 
a disentangled state $\sigma_{AB}$ into an entangled state by local
operations and classical communication. If so, then they can use the so 
obtained entangled state for teleportation. Thus in the
end it would be possible to teleport using disentangled states which
contradicts the previous subsection. Note the last part of the
fundamental law which says ``with no matter how small probability".
This is, of course, very important to stress 
as we have seen that teleportation
is not possible at all with disentangled states. 

In this paper we will work with a more general variant of the above law,
which is more suitable for our purposes. We have seen that
non-local features (i.e. entanglement) cannot be created by
acting locally. This implies that if Alice and Bob share a
certain amount of entanglement (the notion of the amount of entanglement
will be made more precise lateron) initially, they cannot
increase it by only local actions aided with the classical
communication. So we can now restate the fundamental law
in the following, more general, way.

\vspace*{1cm}
\noindent
\fbox{\parbox[b]{16cm}{ \begin{center}
{\bf The fundamental law of quantum 
information processing (2. formulation).}
\end{center}
\vspace*{-0.3cm}
By local operations and classical communication alone, Alice and Bob cannot
increase the total amount of entanglement which they share.}}  
\vspace*{1cm}
\noindent

Note that, contrary to the previous formulation, the addition
``with no matter how small probability" is missing.
This law thus says that the total (or rather, expected) 
entanglement cannot be increased. This still leaves room, 
that with some probability, Alice and Bob can obtain a more entangled
state. Then, however, with some other probability they will 
obtain less entangled states so that on average the 
mean entanglement will not increase.
The above law, it must be stressed, looks deceptively
simple, but we will see that it leads to some profound implications
in quantum information processing. Although it is
derived from considerations of the teleportation protocol, it
nevertheless has much wider consequences. 
For example, we have established 
that if Alice and Bob share disentangled states of the form 
in Eq. (\ref{dis}) then no teleportation is possible. But what 
about the converse: if they share a state not of the form given in 
Eq. (\ref{dis}) can they
always perform teleportation? Namely, even if the state contains
a small amount of entanglement, can that always be used for
teleportation? This amounts to asking whether, given any entangled
state (i.e. a state not of the form in Eq. (\ref{dis})), Alice
and Bob can, with some probability,
obtain the state $(|00\rangle + |11\rangle)/\sqrt{2}$ by
acting only locally and communicating classically. Also we stated that
entanglement cannot increase under local operations, but in
order to check whether it has increased we need some measure
of entanglement.  All these questions 
will be discussed in the following section.  
At the end, we stress that the above law is a 
working assumption and it cannot be proved mathematically. It just so 
happens that by assuming the validity of the fundamental law we can
derive some very useful results, as will be shown 
in the rest of the article.
 
\section{Can we amplify and quantify entanglement?}
In the previous section we have learnt that entanglement is a property that is
essentially different from classical correlations. In particular entanglement 
allows the transmission of an {\em unknown quantum state} using only local 
operations and classical 
communication. Without Alice and Bob sharing one maximally entangled state 
this task can not be achieved perfectly. This impossibility is directly related to the 
fact that it is not possible to create quantum correlations, i.e. entanglement, 
using only local operations and classical communication. This means that if we 
start with a completely uncorrelated state, e.g. a product state,
then local operations and classical communication can only produce a {\em classically 
correlated state}, which is the essence of the fundamental law stated in the
previous section. 
We will now discuss quantum state teleportation again but now not under ideal 
conditions but under circumstances that may occur in an experiment, in particular
under circumstances where decoherence and dissipation are important. This new, 
realistic, situation gives rise to a new idea which is called entanglement purification.

\subsection{Entanglement purification}
In the previous section we have learnt that starting from a product state and 
using only local operations and classical communication, the best we can achieve 
is a classically correlated state, but we will never obtain a state that contains 
any quantum correlations. In particular we will not be able to teleport an unknown 
quantum state if we only share a classically correlated quantum state.

The impossibility of creating entanglement locally poses an important practical 
problem to Alice and Bob when they want to do teleportation in a realistic 
experimental situation. Imagine Alice wants to teleport a quantum state to Bob. 
Furthermore assume that Alice and Bob are really far apart from each other and 
can exchange quantum states only for example through an optical fibre. The fibre, which we will
occasionally call a quantum channel, is really long and it is inevitable that it 
contains faults such as impurities which will disturb the state of a 
photon that we send through the fibre. For teleportation Alice and Bob, need to share 
a maximally entangled state, e.g. a singlet state. However, whenever Alice prepares 
a singlet state on her side and then sends one half of it to Bob the impurities 
in the fibre will disturb the singlet state. Therefore, after the transmission Alice 
and Bob will {\bf not} share a singlet state but some mixed state that is no longer
maximally entangled. If Alice attempts teleportation with this perturbed 
state, Bob will not receive the quantum state Alice tried to send but some perturbed
(and usually mixed) state. Facing this situation, Alice and Bob become quite desperate, 
because they have learnt that it is not possible to create quantum entanglement 
by local operations and classical communication alone. Because Alice and Bob are so
far apart from each other, these are the only operations available to them.
Therefore Alice and Bob conclude that it will be impossible to 'repair' the  state 
they are sharing in order to obtain a perfect singlet between them.  
Luckily Alice and Bob have some friends who are physicists (called say Charles, Gilles, Sandu, 
Benjamin, John and William) and they tell them of their predicament and ask for advice. 
In fact Charles, Gilles, Sandu, Benjamin, John and William confirm that
it is impossible to create entanglement from nothing (i.e. local operations and classical 
communication starting with a product state). However, they inform Alice and Bob that
while it is impossible to create quantum entanglement locally when you have 
no initial entanglement, you can in some sense amplify or, better, concentrate entanglement 
from a source of weakly entangled states to obtain some maximally entangled states
\cite{Bennett96a,Bennett96b,Gisin96,Chiara96,Horodecki97} (this was the more general
formulation of the fundamental law). The purpose of this section is to explain 
briefly two particular implementations (there are too many to discuss all of them) 
of these entanglement purification methods in order 
to convince Alice, Bob and the reader that these methods really work. 

One main difference between the existing purification schemes is their generality, i.e.
whether they can purify an arbitrary quantum state or just certain subclasses such as
pure states. In fact the first purification schemes \cite{Gisin96,Bennett96a}
were not able to purify any arbitrary state. One scheme could purify arbitrary
pure states 
\cite{Bennett96a} (to be described in the following subsection) 
while the other could purify certain special 
classes of mixed state \cite{Gisin96}.
Here we will present a scheme that can purify arbitrary (pure or mixed) bipartite 
states, if these states satisfy one general condition. This condition is expressed 
via the fidelity $F(\rho)$ of the state $\rho$, which is defined as
\begin{equation}
	F(\rho) = \max_{\{all\, max.\, ent. |\psi\rangle\}} \langle \psi| \rho |\psi \rangle \;\; .
	\label{Fidelity}
\end{equation}
In this expression the maximization is taken over all maximally entangled states,
i.e. over all states that one can obtain from a singlet state by local unitary operations.
The scheme we are presenting here requires that the fidelity of the quantum state 
is larger than $0.5$ in order for it to be purifiable. 

\begin{figure}
\centerline{\epsfbox{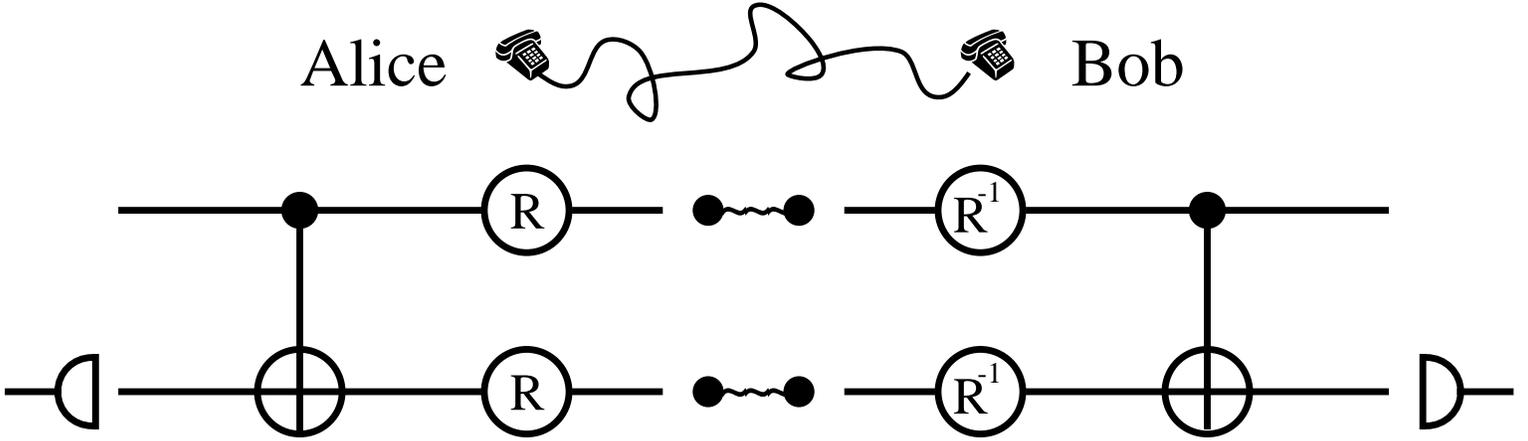}}
\vspace*{1.25cm}
\caption{The quantum network that implements quantum privacy amplification. Alice
and Bob share two pairs of entangled particles. First Alice performs a one bit rotation
${\protect\cal R}$ (given by the R in a circle) which takes 
$\protect{|0\protect\rangle \protect\rightarrow (|0\protect\rangle - i|1\protect\rangle)/\protect\sqrt{2}}$ and
$|1\protect\rangle \protect\rightarrow (|1\protect\rangle - i|0\protect\rangle)/\protect\sqrt{2}$ on her particles, while 
Bob performs the inverse rotation on his side. Then both parties perform a CNOT 
gate on their particles where the first 
pair provides the control bits (signified by the full circle) while the second pair 
provides the target bits (signified by the encircled cross). Finally
Alice and Bob measure the second pair in the $\{0,1\}$ basis. They communicate their
results to each other by classical communication (telephones). If their results
coincide they keep the first pair, otherwise they discard it.}
\label{QPA}
\end{figure}

Although one can perform entanglement purification acting on a single pair of particles 
only \cite{Bennett96a,Gisin96,Vedral97b}, it can be shown that there are states that 
cannot be purified in this way \cite{Popescu98}. Therefore we present a scheme that
acts on two pairs simultaneously. This means that Alice and Bob need to create initially 
two non-maximally entangled pairs of states which they then store. This and the following 
operations are shown in Fig. \ref{QPA}.
Now that Alice and Bob are holding the two pairs, both of them perform two operations. 
First Alice performs a rotation on the two particles she is holding. This rotation
has the effect that
\begin{eqnarray}
	|0\rangle &\rightarrow& \frac{|0\rangle - i |1\rangle}{\sqrt{2}} \\
	|1\rangle &\rightarrow& \frac{|1\rangle - i |0\rangle}{\sqrt{2}} \;\; .
\end{eqnarray}
Bob performs the inverse of this operation on his particles. Subsequently both Alice 
and Bob, perform a controlled NOT (CNOT) gate between the two particles they are holding.
The particle of the first pair serves as the control bit, while the particle of the second
pair serves as the target \cite{Barenco96}. The effect of a CNOT gate is that the second bit 
gets inverted (NOT) when the first bit is in the state $1$ while it remains unaffected when 
the first bit is in the state $0$, i.e.
\begin{eqnarray}
	|0\rangle |0\rangle &\rightarrow& |0\rangle |0\rangle \\
	|0\rangle |1\rangle &\rightarrow& |0\rangle |1\rangle \\
	|1\rangle |0\rangle &\rightarrow& |1\rangle |1\rangle \\
	|1\rangle |1\rangle &\rightarrow& |1\rangle |0\rangle \;\; . 
\end{eqnarray}
The last step in the purification procedure consists of a measurement that both Alice and Bob 
perform on their particle of the second pair. They inform each other about the
measurement result and keep the first pair if their results coincide. Otherwise
they discard both pairs. In each step they therefore discard at least half of the
pairs. From now on we are only interested in those pairs that are not discarded. 
In the Bell basis of Eqs. (\ref{Bell1})-(\ref{Bell}) we define the coefficients
\begin{eqnarray}
	A &=& \langle \Phi^+ | \rho | \Phi^+ \rangle \\
	B &=& \langle \Psi^- | \rho | \Psi^- \rangle \\
	C &=& \langle \Psi^+ | \rho | \Psi^+ \rangle \\
	D &=& \langle \Phi^- | \rho | \Phi^- \rangle \;\;\; .
\end{eqnarray} 
For the state of those pairs that we keep we find that
\begin{eqnarray}
	{\tilde A} &=&  \frac{A^2 + B^2}{N} \label{map1} \\
	{\tilde B} &=&  \frac{2 C D }{N}    \label{map2}\\
	{\tilde C} &=&  \frac{C^2 + D^2}{N} \label{map3}\\
	{\tilde D} &=&  \frac{2 A B}{N}     \label{map4} \;\; .
\end{eqnarray}
Here $N= (A+B)^2 + (C+D)^2$ is the probability that Alice and Bob obtain the same
results in their respective measurements of the second pair, i.e. the probability 
that they keep the first pair of particles. One can quite easily check
that $\{A,B,C,D\} = \{1,0,0,0\}$ is a fixed point of the mapping given in 
Eqs. (\ref{map1} - \ref{map4}) and that for $A>0.5$ one also has ${\tilde A} > 0.5$.
The ambitious reader might want to convince himself numerically that indeed the fixed point 
$\{A,B,C,D\} = \{1,0,0,0\}$ is an attractor for all $A>0.5$, because the analytical
proof of this is quite tricky and not of much interest here. The reader should also
note that the map Eqs. (\ref{map1}) - (\ref{map4}) actually has two fixed points, namely
$\{A,B,C,D\} = \{1,0,0,0\}$ and $\{A,B,C,D\} = \{0,0,1,0\}$. This means that if we want to
know towards which maximally entangled state the procedure will converge, we need
to have some more information about the initial state than just the fidelity according
to Eq. (\ref{Fidelity}). We will not go into further technical details of this purification 
procedure and instead we refer the reader to the literature \cite{Bennett96b,Bennett96c,Deutsch96} 

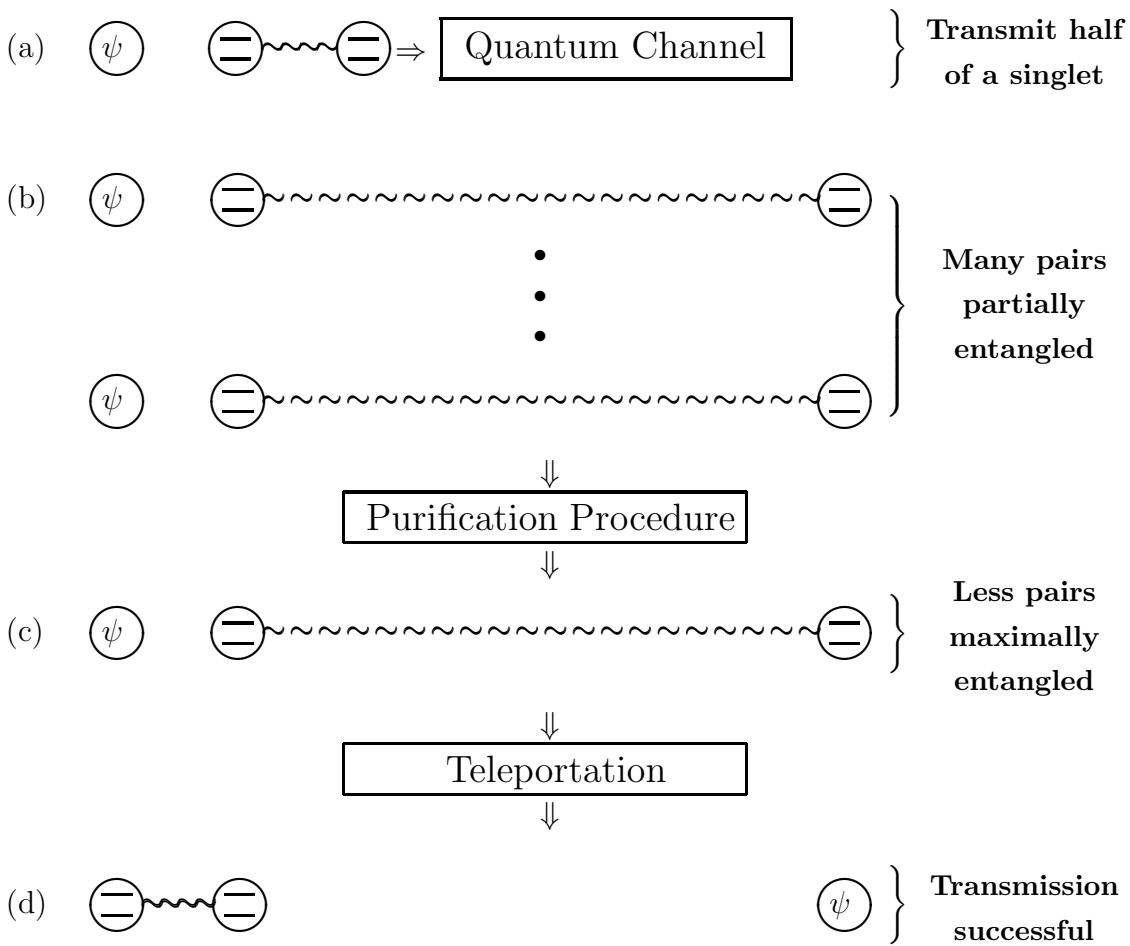
\begin{figure}

\setlength{\unitlength}{1.34mm}
\begin{picture}(75,120)

\thicklines

\put(-8,120){\makebox(0,0)[l]{(a)}}
\put(3,120){\circle{5}}
\put(14.8,120){\circle{5}}
\put(27.5,120){\circle{5}}
\put(1.5,120){\makebox(0,0)[l]{$\small\psi$}}
\put(13.3,121){\line(1,0){3}}
\put(13.3,119){\line(1,0){3}}
\put(26.,121){\line(1,0){3}}
\put(26.,119){\line(1,0){3}}

\put(30.6,118.9){$\Rightarrow$}

\multiput(18.45,120.07)(1.8,0.){4}{\makebox(0,0)[c]{$\bf\sim$}}
\multiput(18.5,119.93)(1.8,0.){4}{\makebox(0,0)[c]{$\bf\sim$}}

\put(35,117){\line(1,0){35}}
\put(69.9,117){\line(0,1){6}}
\put(70,123){\line(-1,0){35}}
\put(35.2,123){\line(0,-1){6}}

\put(37.5,120){\makebox(0,0)[l]{\large Quantum Channel}}

\put(80,120){\makebox(0,0)[c]{$\left. \begin{array}{c} \\[0.1cm]   \end{array} \right\}$ } }

\put(93,120){\makebox(0,0)[c]{$ \small \bf \begin{array}{c} \mbox{ Transmit half}\\[-0.25cm]
\mbox{ of a singlet}\\  \end{array} $ } }



\put(-8,105){\makebox(0,0)[l]{(b)}}

\put(3,105){\circle{5}}
\put(15,105){\circle{5}}
\put(75.2,105){\circle{5}}
\put(1.5,105){\makebox(0,0)[l]{$\small\psi$}}
\put(13.5,106){\line(1,0){3}}
\put(13.5,104){\line(1,0){3}}
\put(73.7,106){\line(1,0){3}}
\put(73.7,104){\line(1,0){3}}

\multiput(18.5,105.05)(2.8,0.){20}{\makebox(0,0)[c]{$\bf\sim$}}
\multiput(18.55,104.95)(2.8,0.){20}{\makebox(0,0)[c]{$\bf\sim$}}

\put(45,99.5){\circle*{1}}
\put(45,95.5){\circle*{1}}
\put(45,91.5){\circle*{1}}

\put(3,85){\circle{5}}
\put(15,85){\circle{5}}
\put(75.2,85){\circle{5}}
\put(1.5,85){\makebox(0,0)[l]{$\small\psi$}}
\put(13.5,86){\line(1,0){3}}
\put(13.5,84){\line(1,0){3}}
\put(73.7,86){\line(1,0){3}}
\put(73.7,84){\line(1,0){3}}

\multiput(18.5,85.05)(2.8,0.){20}{\makebox(0,0)[c]{$\bf\sim$}}
\multiput(18.55,84.95)(2.8,0.){20}{\makebox(0,0)[c]{$\bf\sim$}}

\put(80,94.5){\makebox(0,0)[c]{$\left. \begin{array}{c} \\[0.4cm] \\ \\  \end{array} \right\}$ } }

\put(93,95){\makebox(0,0)[c]{$ \small \bf \begin{array}{c} \mbox{ Many pairs}\\[-0.25cm]
\mbox{ partially}\\[-0.25cm] \mbox{ entangled}\\  \end{array} $ } }

\put(45,77){$\Downarrow$}
\put(25.5,71){\line(1,0){40}}
\put(65.5,71){\line(0,1){5}}
\put(65.5,76){\line(-1,0){40}}
\put(25.5,76){\line(0,-1){5}}
\put(45,68){$\Downarrow$}

\put(46,73.5){\makebox(0,0)[c]{\large Purification Procedure}}


\put(-8,62){\makebox(0,0)[l]{(c)}}

\put(3,62){\circle{5}}
\put(15,62){\circle{5}}
\put(75.2,62){\circle{5}}
\put(1.5,62){\makebox(0,0)[l]{$\small\psi$}}
\put(13.5,63){\line(1,0){3}}
\put(13.5,61){\line(1,0){3}}
\put(73.7,63){\line(1,0){3}}
\put(73.7,61){\line(1,0){3}}

\multiput(18.5,62.05)(2.8,0.){20}{\makebox(0,0)[c]{$\bf\sim$}}
\multiput(18.55,61.95)(2.8,0.){20}{\makebox(0,0)[c]{$\bf\sim$}}

\put(80,62){\makebox(0,0)[c]{$\left. \begin{array}{c} \\[0.1cm]  \end{array} \right \}$ } }

\put(93,62){\makebox(0,0)[c]{$ \small \bf \begin{array}{c} \mbox{ Less pairs}\\[-0.25cm]
\mbox{ maximally}\\[-0.25cm] \mbox{ entangled}\\  \end{array} $ } }

\put(45,52){$\Downarrow$}
\put(25.5,46){\line(1,0){40}}
\put(65.5,46){\line(0,1){5}}
\put(65.5,51){\line(-1,0){40}}
\put(25.5,51){\line(0,-1){5}}
\put(45,43){$\Downarrow$}

\put(46.5,48.2){\makebox(0,0)[c]{\large Teleportation}}


\put(-8,35){\makebox(0,0)[l]{(d)}}

\put(3,35){\circle{5}}
\put(15.2,35){\circle{5}}
\put(75.2,35){\circle{5}}
\put(73.7,35){\makebox(0,0)[l]{$\small\psi$}}
\put(13.7,36){\line(1,0){3}}
\put(13.7,34){\line(1,0){3}}
\put(1.5,36){\line(1,0){3}}
\put(1.5,34){\line(1,0){3}}

\multiput(6.5,35.06)(1.75,0.){4}{\makebox(0,0)[c]{$\bf\sim$}}
\multiput(6.55,34.94)(1.75,0.){4}{\makebox(0,0)[c]{$\bf\sim$}}

\put(80,35){\makebox(0,0)[c]{$\left. \begin{array}{c} \\[0.1cm]   \end{array} \right\}$ } }

\put(93,35){\makebox(0,0)[c]{$ \small \bf \begin{array}{c} \mbox{ Transmission}\\[-0.25cm]
\mbox{ successful}\\  \end{array} $ } }

\end{picture}
\vspace*{.25cm}
\caption{Summary of the teleportation protocol between Alice and Bob in the presence of 
decoherence. (a) Alice (on the left side) holds an unknown quantum state $|\psi\rangle$ 
which she wants to transmit to Bob. Alice creates singlet states and sends one half down 
a noisy channel. (b) She repeats this procedure until Alice and Bob share many partially 
entangled states. (c) Then Alice and Bob apply a local entanglement purification procedure 
to distill a subensemble of pure singlet states. (d) This maximally entangled state can
then be used to teleport the unknown state $|\psi\rangle$ to Bob.}
\label{channel}
\end{figure}

Now let us return to the problem that Alice and Bob wanted to solve, i.e.
to achieve teleportation over a noisy quantum channel. We summarize in 
Fig. \ref{channel} what Alice and Bob have to do to achieve their goal. 
Initially they are given a quantum channel (for example an optical fibre) 
over which they can transmit quantum states. As this quantum channel is not 
perfect, Alice and Bob will end up with a partially entangled state after a 
single use of the fibre. Therefore they repeat the transmission many times 
which gives them many partially entangled pairs of particles. Now they apply 
a purification procedure such as the one described in this section which will 
give them a smaller number of now maximally entangled pairs of particles. With 
these maximally entangled particles Alice and Bob can now teleport an unknown 
quantum state, e.g. $|\psi\rangle$ from Alice to Bob.
Therefore Alice and Bob can achieve perfect transmission of an unknown quantum 
state over a noisy quantum channel. 

\begin{figure}
(a) \centerline{\epsfbox{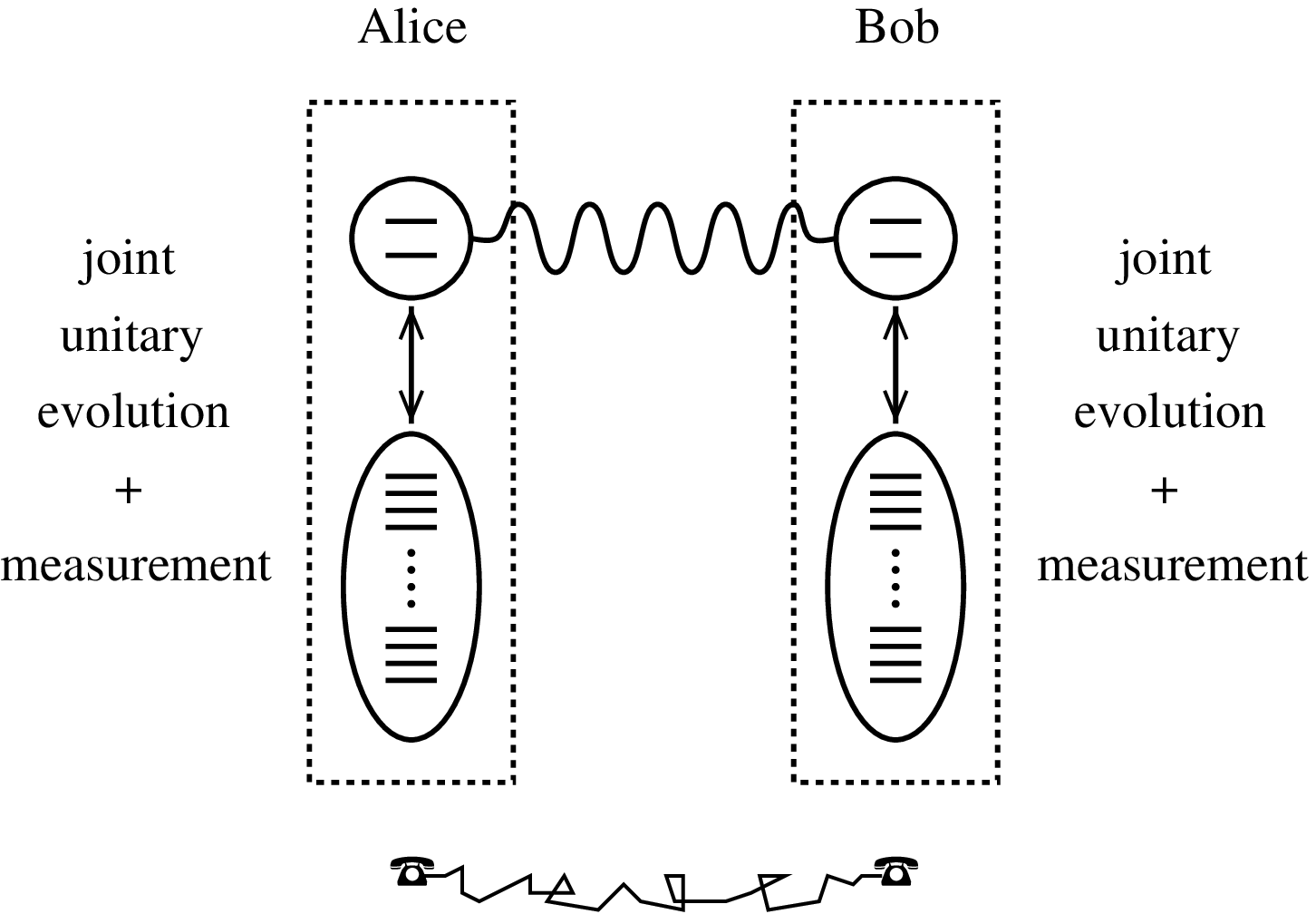}}
\vspace*{2.cm}

(b) \centerline{\epsfbox{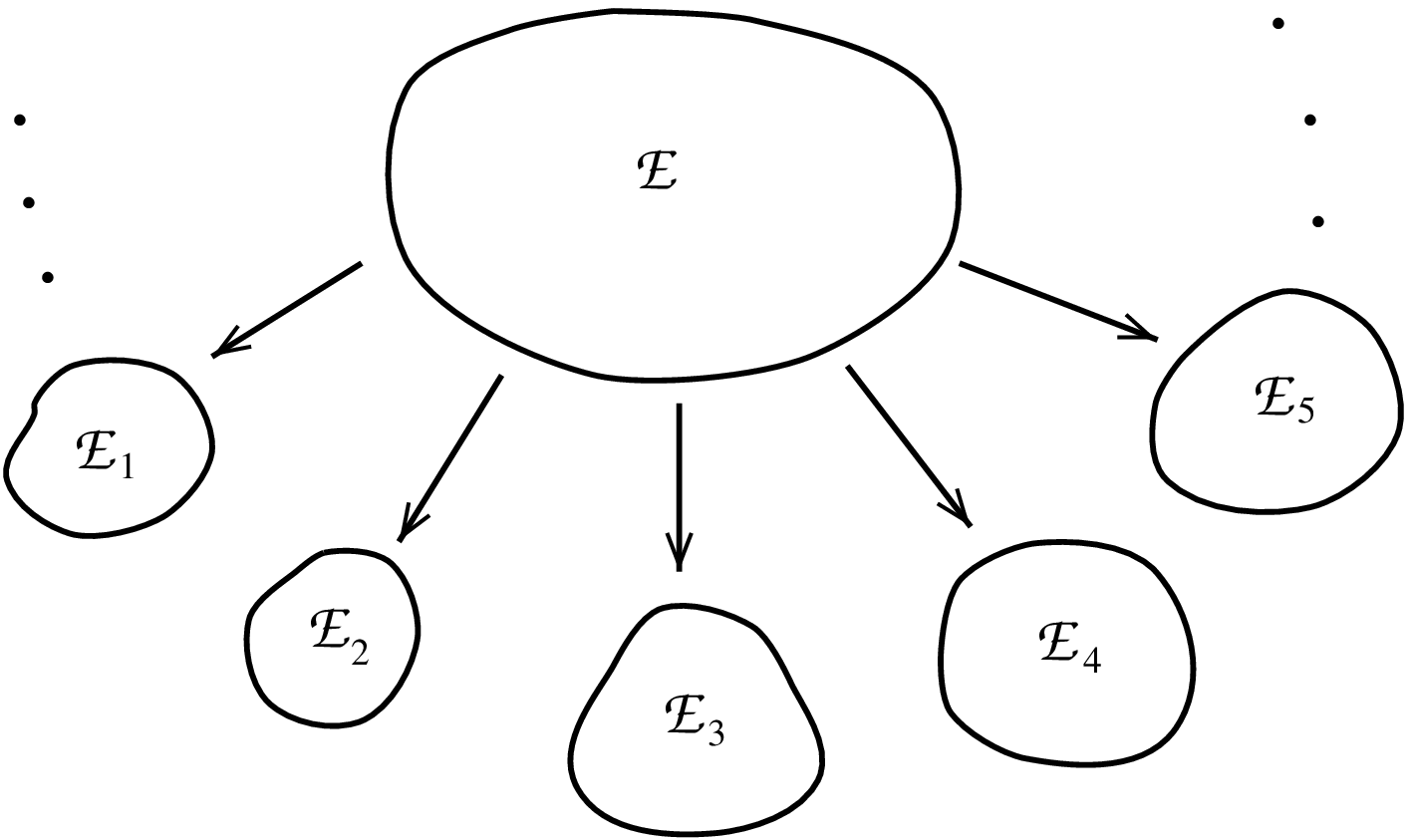}}
\vspace*{1.25cm}
\caption{In quantum state purification procedures three different kinds of operations 
are allowed. In part (a) of this figure the first two are depicted. Alice and 
Bob are allowed to perform any local operation they like. The most general form
is one where Alice adds additional multi-level systems to her particle and then
performs a unitary transformation on the joint system followed by a measurement 
of the additional multi-level system. She can communicate classically with Bob 
about the outcome of her measurement (indicated by the telephones).
The third allowed operation is given in part (b) of the figure.
Using classical communication Alice and Bob can select, based on their
measurement outcomes, subsensembles ${\cal E}_1,\ldots, {\cal E}_n$ 
from the original ensemble ${\cal E}$.
The aim is to obtain at least one subensemble that is in a state having more
entanglement than the original ensemble.}
\label{LGM+CC}
\end{figure}

The main idea of the first two sections of this article are the following.
Entanglement cannot be increased if we are allowed to performed only local 
operations, classical communication and subselection as shown in Fig. \ref{LGM+CC}.
Under all these operations the expected entanglement is non-increasing. This 
implies in particular that, starting from an ensemble in a disentangled state, it is
impossible to obtain entangled states by local operations and classical communication.
However, it does not rule out the possibility that using only local operations
we are able to select from a ensemble described by a partially entangled state a 
subensemble of systems that have higher average entanglement. This is the essence 
of entanglement purification procedures, for which the one outlined here is a 
particular example. Now we review another important purification protocol.

\subsection{Purification of pure states}

The above title is not the most fortunate choice of wording, because 
it might wrongly imply purifying something that is already pure. 
The reader should
remember, however, that the purification means entanglement concentration and
pure states need not be maximally entangled. For example a state 
of the form $a |00\rangle + b |11\rangle$ is not maximally
entangled unless $|a|=|b|=1/\sqrt{2}$. 
In this subsection we consider the following problem first analysed by 
Bennett and coworkers in \cite{Bennett96a}: Alice and Bob
share $n$ entangled qubit pairs, where each pair is prepared in 
the state
\begin{equation}
|\Psi_{AB}\rangle = a |00\rangle + b |11\rangle \;\; ,
\end{equation}
where we take $a, b \in {\cal \bf R}$, and $a^2 + b^2 = 1$. 
How many maximally entangled states can they purify? 
It turns out, that the answer is governed by the von 
Neumann reduced entropy $S_{vN}(\rho_A) \equiv tr \rho_A \ln \rho_A$ 
and is asymptotically given 
by $n\times S_{vN}(\rho_A)=n\times (-a^2\ln a^2 - b^2\ln b^2)$. 
To see why this is so, consider the total state of $n$ pairs given by
\begin{eqnarray}
|\Psi_{AB}^{\otimes n}\rangle & = & (a |00\rangle + b |11\rangle) \otimes (a |00\rangle + b |11\rangle) \otimes \ldots \otimes (a |00\rangle + b |11\rangle)  \nonumber \\
& = & a^{2n} |0000\ldots 00\rangle + a^{2(n-1)} b^2 (|0000 \ldots 11\rangle +\ldots + |1100 \ldots 00\rangle) \nonumber \\ 
& + & \ldots b^{2n} |1111\ldots 11\rangle \; . \label{product}
\end{eqnarray} 
(The convention in the second and the third line is that the states at odd positions
in the large joint ket states belong to Alice and the even states belong
to Bob).
Alice can now perform projections (locally, of course) onto the subspaces 
which have no states $|{1}\rangle$, 2 states $|{1}\rangle$, 4 states 
$|{1}\rangle$, and so on, and communicates her results to Bob. 
The probability of having a successful projection onto a particular
subspace with $2k$ states $|{1}\rangle$ can easily be seen for the 
above equation to be
\begin{equation}
	p_{2k} = a^{2(n-k)}b^{2k} {n \choose k} \; ,
\end{equation} 
which follows directly from Eq. (\ref{product}).
It can be shown that this state can be 
converted into approximately $\ln \left( {n \choose k} \right)$ singlets \cite{Bennett96a}. 
If we assume that the unit of entanglement is given by the entanglement
of the singlet state then the total expected entanglement is seen to be
\begin{equation}
	E = \sum_{k=0}^{n} a^{2(n-k)}b^{2k} {n \choose k} \ln {n \choose k}\; .
\end{equation}
We wish to see how this sum behaves asymptotically as 
$n \rightarrow \infty$. It can be seen easily that the term with the highest 
weight is
\begin{eqnarray}
	E \sim (a^2)^{na^2} (b^2)^{nb^2} {n \choose b^2 n} \ln {n \choose b^2 n} \; ,
\end{eqnarray}
which can, in turn, be simplified using Stirling's approximation to
obtain
\begin{eqnarray}
	E & \sim & e^{-nS_{vN}(\rho_A)} e^{n\ln n - a^2n\ln a^2n -b^2n\ln b^2n} 
	(n\ln n - a^2n\ln a^2n 	-b^2n\ln b^2n) \nonumber \\
	& = & e^{-nS_{vN}(\rho_A)} e^{nS_{vN}(\rho_A)}\times nS_{vN}(\rho_A) = nS_{vN}(\rho_A)\; .
\label{eff}
\end{eqnarray}
This now shows that for pure states the singlet yield of a purification
procedure is determined by the von Neumann reduced entropy. It is
also important to stress that the above procedure is {\em reversible}, 
i.e. starting from $m$ singlets Alice and Bob can locally produce 
a given state $a |0,0\rangle + b |1,1\rangle$ with an asymptotic 
efficiency of $m\ln 2 = n S_{vN} (\rho_A)$. This will be the basis 
of one of the measures of
entanglement introduced by Bennett et al. \cite{Bennett96a}. Of course, 
Alice and Bob cannot do better than this limit, since both 
of them see the initial string of qubits as a classical 
$0,1$ string with the corresponding probabilities 
$a^2$ and $b^2$. This cannot be compressed to more 
than its Shannon entropy $S_{Sh}$ ($S_{Sh} = - a^2 \ln a^2 - b^2 \ln b^2$ 
which in this case coincides with the von Neumann entropy) \cite{Cover91}. However,
another, less technical reason, and more in the spirit of this
article, will be given section 5. 

\section{Entanglement measures}
In the first two sections we have seen that it is possible to concentrate entanglement
using local operations and classical communication. A natural question that arises
in this context is that of the efficiency with which one can perform this concentration.
Given $N$ partially entangled pairs of particles each in the state $\sigma$, how many
maximally entangled pairs can one obtain? This question is basically one about the amount 
of entanglement in a given quantum state. The more entanglement we have initially, the 
more singlet states we will be able to obtain from our supply of non-maximally
entangled states. Of course one could also ask a different question, such as for example:
How much entanglement do we need to create a given quantum state by local operations and 
classical communication alone? This question is somehow the inverse of the question of 
how many singlets we can obtain from a supply of non-maximally entangled states.

All these questions have been worrying physicists in the last two-three years and 
a complete answer is still unknown. The answer to these questions lies in entanglement
measures and in this section we will discuss these entanglement measures a little
bit more. First we will explain conditions every 'decent' measure of entanglement 
should satisfy. After that we will then present some entanglement measures that are
known today. Finally we will compare these different entanglement measures. This 
comparison will tell us something about the way in which the amount of entanglement 
changes under local quantum operations.   

\subsection{Basic properties of entanglement measures}
To determine the basic properties every 'decent' entanglement measure should satisfy we
have to recall what we have learnt in the first two sections of this article. The first
property we realized is that any state of the form Eq. (\ref{dis}), which we call 
separable, does not have any quantum correlations and should therefore be called 
disentangled. This gives rise to our first condition:
\begin{description}
\item[1)] For any separable state $\sigma$ the measure of entanglement should be zero, i.e.
\begin{equation}
	E(\sigma) = 0 \;\; .
	\label{cond1}
\end{equation}
\end{description}
The next condition concerns the behaviour of the entanglement under simple local transformations,
i.e. local unitary transformations. A local unitary transformation simply represents a change
of the basis in which we consider the given entangled state. But a change of basis should not
change the amount of entanglement that is accessible to us, because at any time we could just
reverse the basis change. Therefore in both bases the entanglement should be the same.
\begin{description}
\item[2)] For any state $\sigma$ and any local unitary transformation, i.e. a unitary
transformation of the form $U_A \otimes U_B$, the entanglement remains unchanged. Therefore
\begin{equation}
	E(\sigma) = E(U_A \otimes U_B \sigma U_A^{\dagger} \otimes U_B^{\dagger}) \;\; .
	\label{cond2}
\end{equation}
\end{description}

The third condition is the one that really restricts the class of possible entanglement 
measures. Unfortunately it is usually also the property that is the most difficult 
to prove for potential measures of entanglement. We have seen in section 1 that Alice and 
Bob cannot create entanglement 
from nothing, i.e. using only local operations and classical communication. In section 
2 we have seen that given some initial entanglement we are able to select a subensemble 
of states that have higher entanglement. This can be done using only local operations 
and classical communication. However, what we cannot do is to increase the total amount
of entanglement. We can calculate the total amount of entanglement by summing up the 
entanglement of all systems after we have applied our local operations, classical
communications and subselection. That means that in Fig. \ref{LGM+CC} we take the probability 
$p_i$ that a system will be in particular subensemble ${\cal E}_i$ and multiply it by
the average entanglement of that subensemble. This result we then sum up over all possible 
subensembles. The number we obtain should be smaller than the entanglement of the original 
ensemble.
\begin{description}
\item[3)] Local operations, classical communication and subselection cannot increase the
expected entanglement, i.e. if we start with an ensemble in state $\sigma$ and end up with
probability $p_i$ in subensembles in state $\sigma_i$ then we will have
\begin{equation}
	E(\sigma) \ge  \sum_i p_i E(\sigma_i) \;\; .
	\label{cond3}
\end{equation}
\end{description}
This last condition has an important implication as it tells us something about the 
efficiency of the most general entanglement purification method. To see this we need
to find out what the most efficient purification procedure will look like. Certainly 
it will select one subensemble, which is described by a maximally entangled state. 
As we want to make sure that we have as many pairs as possible in this 
subensemble, we assume that the entanglement in all the other subensembles vanishes. 
Then the probability that we obtain a maximally entangled state from our optimal quantum
state purification procedure is bounded by
\begin{equation}
	p_{singlet} \le \frac{E(\sigma)}{E_{singlet \, state}} \;\; .
	\label{bound}
\end{equation}
The considerations leading to Eq. \ref{bound} show that every entanglement measure
that satisfies the three conditions presented in this section can be used to bound
the efficiency of entanglement purification procedures from above. Before the reader
accepts this statement (s)he should, however, carefully reconsider the above argument.
In fact, we have made a hidden assumption in this argument which is not quite trivial.
We have assumed that the entanglement measures have the property that the entanglement
of two pairs of particles is just the sum of the entanglements of the individual pairs.
This sounds like a reasonable assumption but we should note that the entanglement 
measures that we construct are initially purely  mathematical objects and that we
need to prove that they behave reasonably. Therefore we demand this additivity property as
a fourth condition
\begin{description}
\item[4)] Given two pairs of entangled particles in the total state 
$\sigma = \sigma_1 \otimes \sigma_2$ then we have
\begin{equation}
	E(\sigma) = E(\sigma_1) + E(\sigma_2) \;\; .
\end{equation}
\end{description}

Now we have specified reasonable conditions that any 'decent' measure of entanglement should
satisfy and in the next section we will briefly explain some possible measures of 
entanglement.

\subsection{Three measures of entanglement}
In this subsection we will present three measures of entanglement. 
One of them, the
entropy of entanglement, will be defined only for pure states. 
Nevertheless it is 
of great importance because there are good reasons to accept it 
as the unique measure
entanglement for pure states. Then we will present the 
entanglement of formation
which was the first measure of entanglement for mixed 
states and whose definition
is based on the entropy of entanglement. Finally we 
introduce the relative entropy
of entanglement which was developed from a completely 
different viewpoint. Finally
we will compare the relative entropy of entanglement 
with the entanglement of
formation.

The first measure we are going to discuss 
here is the entropy of entanglement.
It is defined in the following way. Assume 
that Alice and Bob share an entangled 
pair of particles in a state $\sigma$. 
Then if Bob considers his particle alone 
he holds a particle whose state is described 
by the reduced density operator 
$\sigma_B = tr_A\{\sigma\}$. The entropy of 
entanglement is then defined as
the von Neumann entropy of the reduced 
density operator $\sigma_B$, i.e.
\begin{equation}
	E_{vN} = S_{vN}(\sigma_B) = -tr\{\sigma_B \ln \sigma_B \}
	\label{entent}
\end{equation}
One could think that the definition of the entropy 
of entanglement depends on
whether Alice or Bob calculate the entropy of their 
reduced density operator.
However, it can be shown that for a pure state 
$\sigma$ this is not the case, i.e.
both will find the same result. It can be shown 
that this measure of entanglement,
when applied to pure states, satisfies all the 
conditions that we have formulated
in the previous section. This certainly makes 
it a good measure of entanglement.
In fact many people believe that it is the only 
measure of entanglement for pure 
states. Why is that so? In the previous 
section we have learnt that an entanglement
measure provides an upper bound to the 
efficiency of any purification procedure.
For pure states it has been shown that there 
is a purification procedure that
achieves the limit given by the entropy of 
entanglement \cite{Bennett96a}. We reviewed this
procedure in the previous section.
In addition the inverse property has also been shown. Assume that we want 
to create $N$ copies of a quantum state $\sigma$ 
of two particles purely by local operations 
and classical communication. As local operations 
cannot create entanglement,
it will usually be necessary for Alice and Bob to
share some singlets before
they can create the state $\sigma$. How many singlet 
states do they have to share
beforehand? The answer, again, is given by the 
entropy of entanglement, i.e.
to create $N$ copies of a state $\sigma$ of two-particles 
one needs to share
$N\, E(\sigma)$ singlet states beforehand. Therefore we 
have a very interesting
result. The entanglement of pure states can be concentrated 
and subsequently
be diluted again in a reversible fashion. 
One should note, however, that this
result holds only when we have many (actually 
infinitely many) copies of 
entangled pairs at once at our disposal. 
For finite $N$ it is not possible to 
achieve the theoretical limit exactly \cite{Lo97}. This 
observation suggests a close
relationship between entanglement transformations 
of pure states and thermodynamics.
We will see in the following to what extent this 
relationship extends to mixed entangled
states.

We will now generalize the entropy of entanglement to mixed states. It will turn
out that for mixed states there is not one unique measure of entanglement 
but that there are several different measures of entanglement. 

How can we define a measure of entanglement for mixed states? As we now have agreed 
that the entropy of entanglement is a good measure of entanglement for pure states,
it is natural to reduce the definition of mixed state entanglement to that of pure
state entanglement. One way of doing that is to consider the amount of entanglement
that we have to invest to create a given quantum state $\sigma$ of a pair of particles.
By creating the state we mean that we represent the state $\sigma$ by a statistical
mixture of pure states. It is important in this representation that we do not restrict
ourselves to pure states that are orthonormal. If we want to attribute an amount of
entanglement to the state $\sigma$ in this way then this should be the smallest amount
of entanglement that is required to produce the state $\sigma$ by mixing pure states
together. If we measure the entanglement of pure states by the entropy of entanglement,
then we can define the entanglement of formation by
\begin{equation}
	E_F(\sigma) = \min_{\sigma = \sum_i p_i |\psi_i\rangle\langle \psi_i|}
		      \sum_i p_i E_{vN}(|\psi_i\rangle\langle \psi_i|) \;\; .
	\label{formation}
\end{equation}  
The minimization in Eq. (\ref{formation}) is taken over all possible decompositions
of the density operator $\sigma$ into pure states $|\psi\rangle$. In general, this
minimization is extremely difficult to perform. Luckily for pairs of two-level systems
one can solve the minimization analytically and write down a closed expression
for the entanglement of formation which can be written entirely in terms of the 
density operator $\sigma$ and does not need any reference to the states of the
optimal decomposition. In addition the optimal decomposition of $\sigma$ can be
constructed for pairs of two-level systems. To ensure that Eq. (\ref{formation}) really
defines a measure
of entanglement, one has to show that it satisfies the four conditions we have 
stated in the previous section. The first three conditions can actually be proven 
analytically (we do not present the proof here) while the fourth condition (the 
additivity of the entanglement) has so far only been confirmed numerically. 
Nevertheless the entanglement of formation is a very important measure of entanglement
especially because there exists a closed analytical form for it \cite{Wootters98}. 

As the entanglement of formation is a measure of entanglement it represents an upper
bound on the efficiency of purification procedures. However, in addition it 
also gives the amount of entanglement that has to be used to create a given quantum
state. This definition of the entanglement of formation alone guarantees already
that it will be an upper bound on the efficiency of entanglement purification.
This can be seen easily, because if there would be a purification procedure that
produces, from $N$ pairs in state $\sigma$, more entanglement than 
$N\,E_F(\sigma)$ then we would be able to use this entanglement to create more
than $N$ pairs in the state $\sigma$. Then we could repeat the purification procedure
and we would get even more entanglement out. This would imply that we would be able
to generate arbitrarily large amounts of entanglement by purely local operations
and classical communication. This is impossible and therefore the entanglement of 
formation is an upper bound on the efficiency of entanglement purification. What
is much more difficult to see is whether this upper bound can actually be achieved 
by any entanglement purification procedure. On the one hand we have seen that for
pure states it is possible to achieve the efficiency bound given by the entropy of 
entanglement. On the other hand for mixed states the situation is much more complicated 
because we have the additional statistical uncertainty in the mixed state. We would expect
that we have to make local measurements in order to remove this statistical uncertainty and
these measurements would then destroy some of the entanglement. On the other hand
we have seen that in the pure state case we could recover all the entanglement despite
the application of measurements. This question
was unresolved for some time and it was possible to solve it when yet another 
measure of entanglement, the relative entropy of entanglement, was discovered. 

\begin{figure}
\centerline{\epsfbox{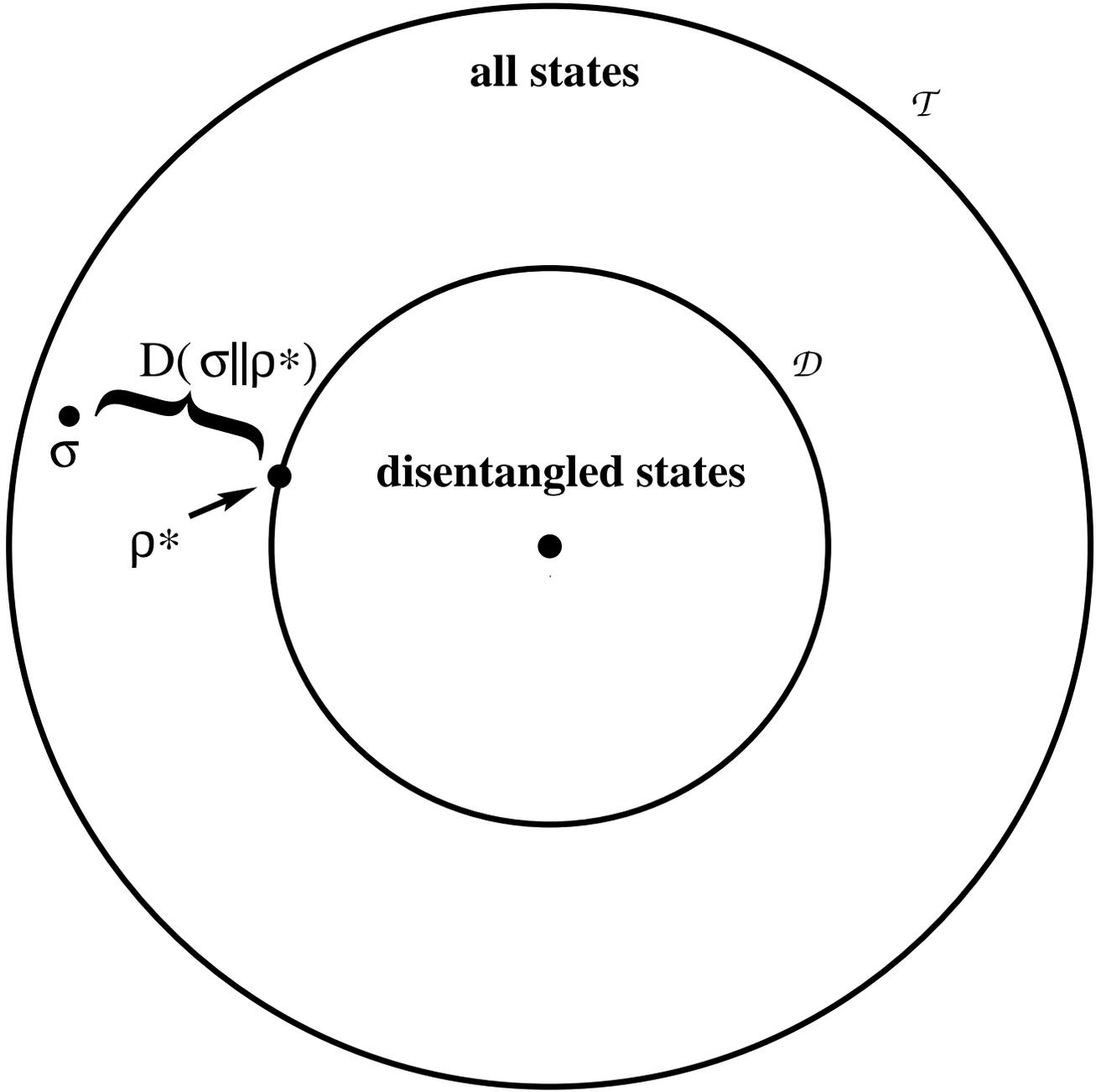}}
\vspace*{1.25cm}
\caption{A geometric way to quantify entanglement. The set of all density 
matrices ${\cal T}$ is represented 
by the outer circle. Its subset of disentangled (separable) states ${\cal D}$, is 
represented by the inner circle. A state $\sigma$ belongs to the 
entangled states, and $\rho^*$ is the disentangled state that minimizes 
the distance $D(\sigma||\rho)$. This minimal distance can be defined as
the amount of entanglement in $\sigma$.}
\label{relent}
\end{figure}

The relative entropy of entanglement has been introduced in a different way than
the two entanglement measures presented above \cite{Vedral97a,VP98a}. The basic ideas in the relative 
entropy of entanglement are based on distinguishability and geometrical distance.
The idea is to compare a given quantum state $\sigma$ of a pair of particles with
disentangled states. A canonical disentangled state that one can form from $\sigma$
is the state $\sigma_A \otimes \sigma_B$ where $\sigma_A$ ($\sigma_B$) is the reduced
density operator that Alice (Bob) are observing. Now one could try to define the 
entanglement of $\sigma$ by any distance between $\sigma$ and 
$\sigma_A \otimes \sigma_B$. The larger the distance the larger is the entanglement of 
$\sigma$. 
Unfortunately it is not quite so easy to make an entanglement measure. The problem is that 
we have picked a particular (although natural) disentangled state. Under a
purification procedure this product state $\sigma_A \otimes \sigma_B$ can be turned into a 
sum of product states, i.e. a classically correlated state.
But what we know for sure is that under any purification procedure a separable state of
the form Eq. (\ref{dis})
will be turned into a separable state. Therefore it would be much more natural to compare
a given state $\sigma$ to all separable states and then find that separable state that
is closest to $\sigma$. This idea is presented in Fig. \ref{relent} and can be written in a 
formal
way as
\begin{equation}
	E_{RE}(\sigma) = \min_{\rho\in{\cal D}} D(\sigma||\rho) \;\; .
	\label{relent1}
\end{equation}
Here the ${\cal D}$ denotes the set of all separable states and $D$ can be any function that 
describes a measure of separation between two density operators. Of course, not all distance
measures will generate a 'decent' measure of entanglement that satisfies all the conditions 
that we demand from an entanglement measure. Fortunately, it is possible to find some distances
$D$ that generate 'decent' measures of entanglement and a particularly nice one is the 
relative entropy which is defined as
\begin{equation}
	S(\sigma||\rho) = tr\{ \sigma \ln\sigma - \sigma \ln\rho \} \;\; .
	\label{relative}
\end{equation}
The relative entropy is a slightly peculiar function and is in fact not really a distance in the
mathematical sense because it is not even symmetric. Nevertheless it can be proven that Eq. (\ref{relent1})
together with the relative entropy of Eq. (\ref{relative}) generates a measure of entanglement
that satisfies all the conditions we were asking for in the previous section. It should be said here
that the additivity of the relative entropy of entanglement has only been confirmed numerically as 
for the entanglement of formation. All other properties can be proven analytically and it should also
be noted that for pure states the relative entropy of entanglement reduces to the entropy of 
entanglement which is of course a very satisfying property. 

\begin{figure}
\centerline{\epsfbox{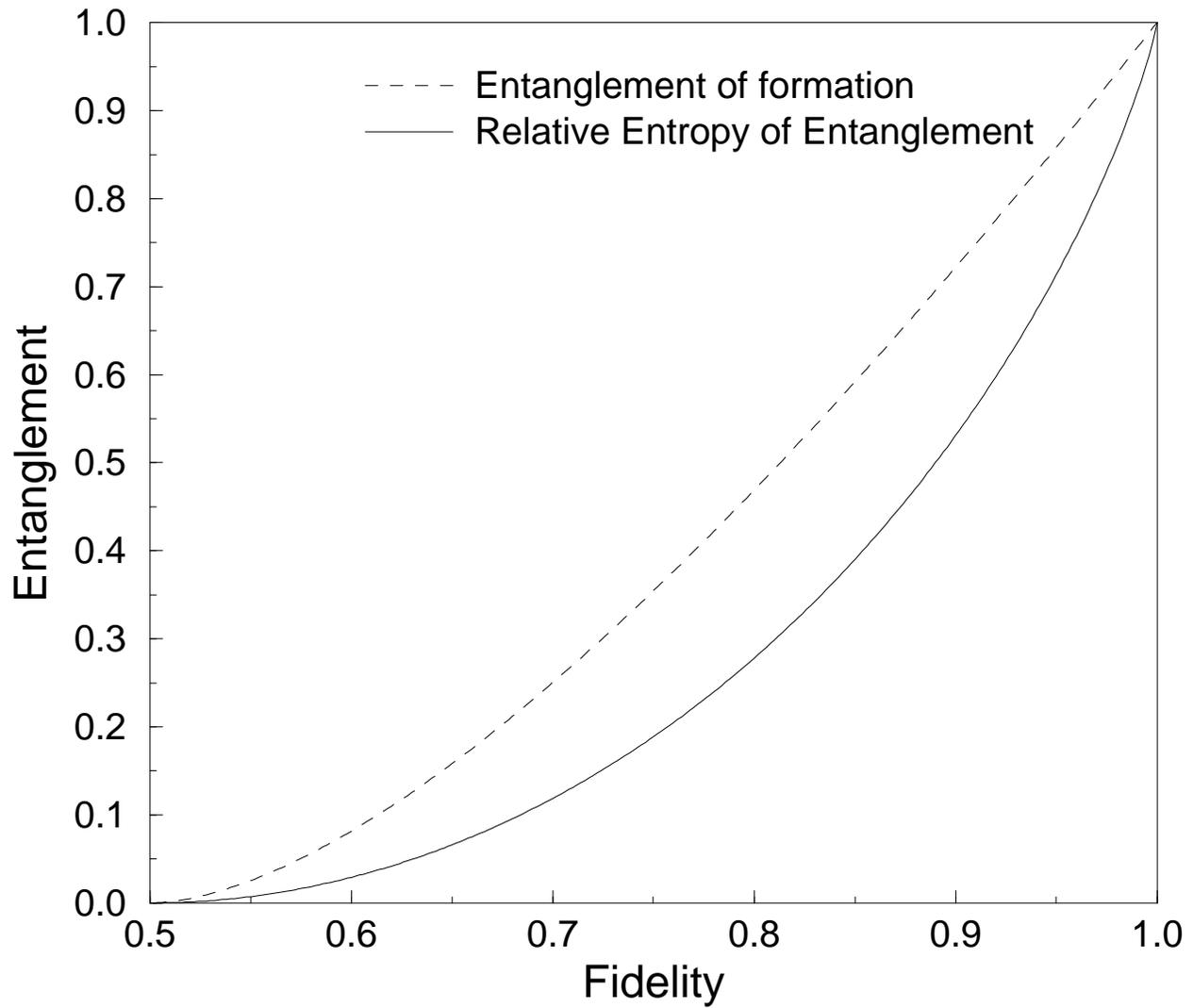}}
\caption{Comparison of the entanglement of formation with the relative entropy
of entanglement for Werner states with fidelity $F$. The relative entropy of
entanglement is always smaller than the entanglement of formation. This proves 
that in general entanglement is destroyed by local operations. }
\label{newbound}
\end{figure}

But why does the relative entropy of entanglement answer the question whether the upper bound
on the efficiency of entanglement purification procedures that we found from the entanglement
of formation can actually be achieved or not? The answer comes from a direct comparison of
the two measures of entanglement for a particular kind of state. These, called Werner states, 
are defined as 
\begin{equation}
	\rho_F = F | \psi^{-}\rangle \langle \psi^{-} | + \frac{1-F}{3} (| \psi^{+}\rangle \langle \psi^{+} |
+ | \phi^{-}\rangle \langle \phi^{-} | + | \phi^{+}\rangle \langle \phi^{+} | ) 
	\label{werner}
\end{equation}
where we have used the Bell basis defined in Eq. (\ref{Bell1}-\ref{Bell}). 
The parameter $F$ is the fidelity of the
Werner state and lies in the interval $[\frac{1}{4},1]$. For Werner states it is possible
to calculate both the entanglement of formation and the relative entropy of entanglement
analytically. In Fig. \ref{newbound} the entanglement of the Werner states with fidelity $F$
is plotted for both entanglement measures. One can clearly see that the relative entropy of
entanglement is smaller than the entanglement of formation. But we know that the relative entropy
of entanglement, because it is an entanglement measure, is an upper bound on the efficiency 
of any entanglement purification procedure too. Therefore we reach the following very interesting 
conclusion. Assume we are given a certain amount of entanglement that we invest in the most optimal
way to create by local means some mixed quantum states $\sigma$ of pairs of two-level systems. 
How many pairs in the state $\sigma$ we can produce is determined by the entanglement of formation. 
Now we try to recover this entanglement by an entanglement purification method whose efficiency
is certainly bounded from above by the relative entropy of entanglement. The conclusion is that
the amount of entanglement that we can recover is always smaller than the amount of entanglement 
that we originally invested. Therefore we arrive at an irreversible 
process, in stark contrast to the pure state case where 
we were able to recover all the invested entanglement by a 
purification procedure. This result again sheds some light 
on the connection between entanglement 
manipulations and thermodynamics and in the next 
section we will elaborate on this connection further.

\section{Thermodynamics of entanglement}

Here we would like to elucidate further the fundamental law
of quantum information processing by comparing it 
to the Second Law of Thermodynamics. The reader should not
be surprised that there are connections between the
two. First of all, both laws can be expressed mathematically
by using an entropic quantity. The second law says that 
thermodynamical entropy cannot decrease in an isolated
system. The fundamental law of quantum information processing, on the other hand, states
that entanglement cannot be increased by local operations.
Thus both of the laws serve to prohibit certain
types of processes which are impossible in nature (this 
analogy was first emphasised by Popescu and Rohrlich in 
\cite{Popescu97}, but also see \cite{Horodecki97a,VP98a}).
The rest of the section shows the two principles in
action by solving two simple, but important problems.

\subsection{Reversible and irreversible processes}

We begin by stating more formally a form of the
Second Law of thermodynamics. This form is due to Clausius, but it
is completely analogous to the no increase of entropy
statement we gave above. In particular it will be
more useful for what we are about to investigate.

\vspace*{1cm}
\noindent
\fbox{\parbox[b]{16cm}{ {\bf The Second Law of thermodynamics 
(Clausius).}
There exists no thermodynamic process the {\em sole} effect of which is
to extract a quantity of heat from the colder of two reservoirs and 
deliver it to the hotter of the two reservoirs.}} 
\vspace*{1cm}

Suppose now that we have a thermodynamical system. We want to
invest some heat into it so that at the end our system does
as much work as possible with this heat input. The 
efficiency is therefore defined as
\begin{equation}
\eta  = \frac{W_{\mbox{out}}}{Q_{\mbox{in}}} \;\; .
\end{equation} 
Now it is a well known fact that the above efficiency is maximized
if we have a reversible process (simply because an irreversible
process wastes useful work on friction or some other
lossy mechanism). In fact, we know the efficiency of one
such process, called the Carnot cycle. With the Second Law on
our mind, we can now prove that no other process can perform
better than the Carnot cycle. This boils down to the
fact that we only need to prove that 
no other reversible process performs better than the 
Carnot cycle. The argument for this can be found in any
undergraduate book on Thermodynamics and briefly runs as follows
(again reductio ad absurdum).
The Carnot engine takes some  heat input from a hotter 
reservoir, does some work and delivers an amount of heat 
to the colder reservoir. Suppose that there is a 
better engine, E, that is operating between the same two
reservoirs (we have to be fair when comparing the
efficiency). Suppose also that we run this better machine
backwards (as a refrigerator): we would do some work on it,
and it would take a quantity of heat from the cold reservoir
and bring some heat to the hot reservoir. For simplicity we
assume that the work done by a Carnot engine is the same as the
work that E needs to run in reverse (this can always be arranged and
we lose nothing in generality). 
Then we look at the two machines together, which is just another
thermodynamical process: they extract a quantity of
heat from the colder reservoir and deliver it to the hot reservoir
with all other things being equal. But this contradicts the 
Second Law, and therefore no machine is more efficient
then the Carnot engine.

In the previous section we have learnt
about the purification scheme of Bennett et al \cite{Bennett96a}
for pure states. Efficiency of any scheme was defined
as the number of maximally entangled states we can obtain
from a given $N$ pairs in some initial state, divided by $N$. 
This scheme is in addition reversible,
and we would suppose, guided by the above thermodynamic
argument, that no other reversible purification scheme 
could do better than the Bennett et al. Suppose that 
there is a more efficient (reversible) process.
Now Alice and Bob start from a certain number N of maximally
entangled pairs. They apply a reverse of the scheme of Bennett et al \cite{Bennett96a}
to get a certain number of less entangled states. But
then they can run the more efficient purification to get M
maximally entangled states out. However, since the second 
purification is more efficient than the first one, then we have that
$M>N$. So, locally
Alice and Bob can increase entanglement, which contradicts
the fundamental law of quantum information processing. 
We have to stress that as far as the mixed states
are concerned there are no results regarding the
best purification scheme, and it is not completely
understood whether the same strategy as above
could be applied (for more discussion see \cite{VP98a}).

In any case, the above reasoning shows that the conceptual ideas 
behind the Second law and the fundamental law
are similar in nature. Next we show another 
attractive application of the fundamental law. We
return to the question at the beginning of the
article that started the whole discussion:
can Alice teleport to Bob as many qubits as she likes 
using only one entangled pair shared between them?

\subsection{What can we learn from the non-increase of 
entanglement under local operations?}

If the scheme that we are proposing could be utilized
then it would be of great technological advantage,
because to create and maintain entangled qubits
is at present very hard. If a single maximally entangled pair could
transfer a large amount of information (i.e. 
teleport a number of qubits), then this would 
be very useful. However, there is no free lunch.
In the same way that we cannot have an unlimited
amount of useful work and no heat dissipation,
we cannot have arbitrarily many teleportations 
with a single maximally entangled pair. In fact, we can
prove a much stronger statement: in order to
teleport $N$ qubits Alice and Bob need to share
$N$ maximally entangled pairs! 

In order to prove this we need to understand another
simple concept from quantum mechanics. Namely, 
if we can teleport a pure unknown quantum state
then we can teleport an unknown mixed quantum
state (this is obvious since a mixed state is
just a combination of pure states). But now
comes a crucial result: every mixed state 
of a single qubit can be thought of as a part of
a pure state of two {\em entangled} qubits (this result is more
general, and applies to any quantum state of any quantum system, 
but we do not need the generalization here).
Suppose that we have a single qubit in a state 
\begin{equation}
 \rho = a^2 |0\rangle\langle 0| + b^2 |1\rangle\langle 1| \label{rho} \; .
\end{equation}
This single qubit can then be viewed as a part of a pair of qubits in state
\begin{equation}
 |\psi\rangle = a |00\rangle + b |11\rangle  \label{psi}\; .
\end{equation} 
One obtains Eq. (\ref{rho}) from Eq. (\ref{psi}) simply by taking the 
partial trace over the second particle. Bearing this in
mind we now envisage the following teleportation 
protocol. Alice and Bob share a maximally entangled pair,
and in addition Bob has a qubit prepared in some 
state, say $|0\rangle$. Alice than receives a qubit to 
teleport in a general (to her unknown) state $\rho$. 
After the teleportation 
we want Bob's extra qubit to be in the state $\rho$ and
the maximally entangled pair to stay intact (or at least
not to be completely destroyed). This is shown in Fig. \ref{destroy1}.

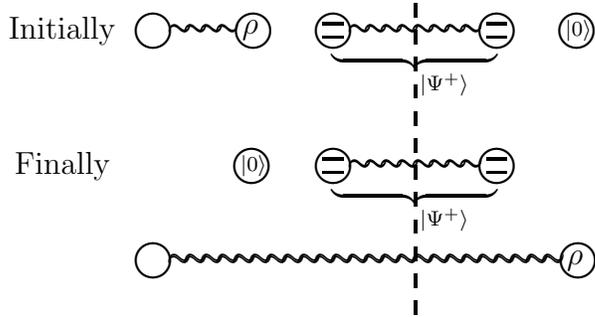
\begin{figure}[hbt]

\setlength{\unitlength}{0.9mm}
\begin{picture}(75,48)

\thicklines

\put(15,42){\makebox(0,0)[c]{Initially}}
\put(28.4,42){\circle{5}}
\put(43.2,42){\circle{5}}
\put(55,42){\circle{5}}
\put(79.2,42){\circle{5}}
\put(91,42){\circle{5}}

\put(53.5,43){\line(1,0){3}}
\put(53.5,41){\line(1,0){3}}
\put(77.7,43){\line(1,0){3}}
\put(77.7,41){\line(1,0){3}}

\put(41.6,42){\makebox(0,0)[l]{$\small\rho$}}
\scriptsize
\put(89.3,42){\makebox(0,0)[l]{$ |0\rangle$}}
\normalsize

\put(54.8,37.5){\makebox(0,0)[l]{$\underbrace{\hspace{2.2cm}}$ }}
\scriptsize
\put(72,34){\makebox(0,0)[c]{$|\Psi^{+}\rangle$ }}
\normalsize

\multiput(58.95,42.1)(2.35,0.){8}{\makebox(0,0)[c]{$\bf\sim$}}
\multiput(58.9,41.9)(2.35,0.){8}{\makebox(0,0)[c]{$\bf\sim$}}

\multiput(32.35,42.1)(2.35,0.){4}{\makebox(0,0)[c]{$\bf\sim$}}
\multiput(32.3,41.9)(2.35,0.){4}{\makebox(0,0)[c]{$\bf\sim$}}

\multiput(67.1,46)(0,-4){12}{\line(0,-1){2}}
\multiput(67.3,46)(0,-4){12}{\line(0,-1){2}}


\put(15,22){\makebox(0,0)[c]{Finally}}
\put(28.4,8){\circle{5}}
\put(43,22){\circle{5}}
\put(55,22){\circle{5}}
\put(79.2,22){\circle{5}}
\put(91.2,8){\circle{5}}

\put(53.5,23){\line(1,0){3}}
\put(53.5,21){\line(1,0){3}}
\put(77.7,23){\line(1,0){3}}
\put(77.7,21){\line(1,0){3}}

\put(89.6,8){\makebox(0,0)[l]{$\small\rho$}}
\scriptsize
\put(41.35,22){\makebox(0,0)[l]{$ |0\rangle$}}
\normalsize

\put(54.8,17.5){\makebox(0,0)[l]{$\underbrace{\hspace{2.2cm}}$ }}
\scriptsize
\put(72,14){\makebox(0,0)[c]{$|\Psi^{+}\rangle$ }}
\normalsize

\multiput(58.95,22.1)(2.35,0.){8}{\makebox(0,0)[c]{$\bf\sim$}}
\multiput(58.9,21.9)(2.35,0.){8}{\makebox(0,0)[c]{$\bf\sim$}}

\multiput(32.3,8.1)(2.4,0.){24}{\makebox(0,0)[c]{$\bf\sim$}}
\multiput(32.25,7.9)(2.4,0.){24}{\makebox(0,0)[c]{$\bf\sim$}}

\end{picture}
\caption{A diagramatical proof that the teleportation protocol in 
Fig. (\protect\ref{destroy1}) is impossible. Alice is on the left of the
dashed line, Bob on the right. Initially Alice is holding a mixed state
$\rho$ and Bob a particle in state $|0\rangle$. In addition Alice and 
Bob share a pair of maximally entangled particles in state $|\psi^+\rangle$. 
The particle in the
mixed state $\rho$ that Alice is holding can be part of a pair of entangled 
particles. The aim is that finally, after the teleportation Bob holds the state 
$\rho$ and Alice and Bob still have their two particles in a maximally
entangled state $|\psi^+\rangle$. However, not only the state $\rho$ will
be transferred to Bob but also its entanglement with other particles.
Therefore after the envisaged teleportation Alice and Bob would be 
sharing more entanglement than initially. This contradicts the 
fundamental law of quantum information processing that entanglement 
cannot be increased.}
\label{destroy2}
\end{figure}

Now we wish to prove this protocol impossible--entanglement
simply has to be completely destroyed at the end. 
Suppose it is not, i.e. suppose that the above teleportation is 
possible. Then Alice can teleport any unknown (mixed) state to Bob
using this protocol.
But this mixed state can arise from an entangled state
where the second qubit (the one to be traced out) is on Alice's
side. So initially Alice and Bob share one entangled pair, but
after the teleportation they have increased their entanglement as
in Fig. \ref{destroy2}. Since the initial state can be maximally 
mixed state ($a=b=1/\sqrt{2}$) the final entanglement can
grow to be twice the maximally entangled state. But, as
this would violate the fundamental law of quantum information processing it is impossible 
and the initial maximally entangled pair has to be destroyed.
In fact, this argument shows that it has to be destroyed 
completely. Thus we see that a simple application of the
fundamental law can be used to rule out a whole class of
impossible teleportation protocols. Otherwise every teleportation protocol
would have to be checked separately and this would be a very hard problem.

\section{Conclusions}

Let us briefly recapitulate what we have learnt. Quantum
teleportation is a procedure whereby an unknown state
of a quantum system is transferred from a particle at a place 
A to a particle at a place B. The whole protocol uses only local
operations and classical communication between A and B. In addition,
A and B have to share a maximally entangled state. Entanglement
is central for the whole teleportation: if that
state is not maximally entangled then teleportation is
less efficient and if the state is disentangled (and only
classically correlated) then teleportation is impossible.
We have then derived a fundamental law of quantum information
processing which stipulated that entanglement cannot be increased
by local operations and classical communication only.  
This law was then investigated in the light of purification
procedures: local protocols for increasing entanglement
of a subensemble of particles. We discussed bounds on the
efficiency of such protocols and emphasised the links
between this kind of physics and the theory of thermodynamics.
This lead us to formulate various measures of entanglement
for general mixed states of two quantum bits. At the
end we returned to the problem of teleportation, asking
how many entangled pairs we need in order to teleport N qubits.
Using the fundamental law of quantum information
processing we offered an elegant argument for
needing N maximally entangled pairs for teleporting N qubits,
a pair per qubit. 

The analogy between thermodynamics and
quantum information theory might be deeper, but this at
present remains unknown. Quantum information theory is still 
at a very early stage of development 
and, although there are already some
extraordinary results, a number of areas is still untouched.
In particular the status of what we called the fundamental 
law is unclear. First and foremost, it is not known how it relates to other
results in the field, such as, for example, the no 
cloning theorem \cite{Wootters82} which states that an unknown quantum state
cannot be duplicated by a physical process. We hope that research in this area
will prove fruitful in establishing a deeper 
symbiotic relationship between information 
theory, quantum physics and thermodynamics.
Quantum theory has had a huge input into information
theory and thermodynamics over the past few decades. 
Perhaps by turning this 
around we can learn much more about
quantum theory by using information-theoretic and
thermodynamic concepts. Ultimately, this approach
might solve some long standing and difficult problems 
in modern physics, such as the measurement problem and the
arrow of time problem. This is exactly what was envisaged
more that 60 years ago in a statement attributed to Einstein: 
'The solution of the problems of
quantum mechanics will be thermodynamical in nature' \cite{Einstein}.

\section{Acknowledgements}
The authors would like to thank Susana F. Huelga and Peter L. Knight
for critical reading of the manuscript. This work was supported in
part by Elsag-Bailey, the UK Engineering and Physical Sciences
Research Council (EPSRC) and the European TMR Research
Network ERBFMRXCT960066 and the European TMR Research Network
ERBFMRXCT960087.

\newpage

\begin{center}
	Biographies
\end{center}

Martin Plenio studied in G{\"o}ttingen (Germany) where he obtained both his Diploma (1992) and
his PhD (1994) in Theoretical Physics. His main research area at that time was 
Quantum Optics and in particular the properties of single quantum systems such
as single trapped ions irradiated by laser light. After his PhD he joined 
the Theoretical Quantum Optics group at Imperial College as a postdoc. It was here 
that he started to become interested in quantum computing, quantum
communication and quantum information theory. Since January 1998 he is now
a lecturer in the Optics Section of Imperial College.\\[2.cm]

Vlatko Vedral obtained both his first degree (1995) and PhD (1998) in
Theoretical Physics from Imperial College. He is now an 
Elsag-Bailey Postdoctoral Research Fellow
at the Center for Quantum Computing in Oxford. From October 1998 he
will take up a Junior Research Fellowship at Merton College in
Oxford. His main research interests are in connections between 
information theory and quantum mechanics, 
including quantum computing, error correction and quantum 
theory of communication.

\newpage

\begin{figure}[hbt]

\setlength{\unitlength}{0.9mm}
\begin{picture}(75,48)

\thicklines

\put(0,42){\makebox(0,0)[l]{(a)}}
\put(13,42){\circle{5}}
\put(25,42){\circle{5}}
\put(49.2,42){\circle{5}}
\put(11.5,42){\makebox(0,0)[l]{$\small\psi$}}
\put(23.5,43){\line(1,0){3}}
\put(23.5,41){\line(1,0){3}}
\put(47.7,43){\line(1,0){3}}
\put(47.7,41){\line(1,0){3}}

\put(25,37.5){\makebox(0,0)[l]{$\underbrace{\hspace{2.cm}}$ }}
\scriptsize
\put(37,34){\makebox(0,0)[c]{$|\Psi^{+}\rangle$ }}
\normalsize

\multiput(28.95,42.1)(2.35,0.){8}{\makebox(0,0)[c]{$\bf\sim$}}
\multiput(28.9,41.9)(2.35,0.){8}{\makebox(0,0)[c]{$\bf\sim$}}

\scriptsize
\put(60,42){\makebox(0,0)[l]{$(\alpha|0\rangle + \beta |1\rangle)(|00\rangle + |11\rangle)/\sqrt{2}$ }}
\normalsize


\put(0,12){\makebox(0,0)[l]{(b)}}
\put(13,12){\circle{5}}
\put(25,12){\circle{5}}
\put(49.2,12){\circle{5}}
\put(11.5,12){\makebox(0,0)[l]{$\small\psi$}}
\put(23.5,13){\line(1,0){3}}
\put(23.5,11){\line(1,0){3}}
\put(47.7,13){\line(1,0){3}}
\put(47.7,11){\line(1,0){3}}

\multiput(9,8)(4,0){5}{\line(1,0){2}}
\multiput(9,8.2)(4,0){5}{\line(1,0){2}}
\multiput(29,8)(0,4){2}{\line(0,1){2}}
\multiput(28.8,8)(0,4){2}{\line(0,1){2}}
\multiput(29,16)(-4,0){5}{\line(-1,0){2}}
\multiput(29,15.8)(-4,0){5}{\line(-1,0){2}}
\multiput(9,16)(0,-4){2}{\line(0,-1){2}}
\multiput(8.8,16)(0,-4){2}{\line(0,-1){2}}

\small
\put(9,6){\makebox(0,0)[l]{Measurement}}
\scriptsize
\put(9,1){\makebox(0,0)[l]{$\{|\Psi^{\pm}\rangle, |\Phi^{\pm}\rangle$\}  }}
\normalsize

\multiput(28.95,12.1)(2.35,0.){8}{\makebox(0,0)[c]{$\bf\sim$}}
\multiput(28.9,11.9)(2.35,0.){8}{\makebox(0,0)[c]{$\bf\sim$}}

\scriptsize
\put(60,19){\makebox(0,0)[l]{$\;\;\,  \frac{1}{2}|\Psi^{+}\rangle (\alpha|0\rangle + \beta |1\rangle)$ }}
\put(60,14.5){\makebox(0,0)[l]{$+ \frac{1}{2}|\Psi^{-}\rangle (\alpha|0\rangle - \beta |1\rangle)$ }}
\put(60,9.5){\makebox(0,0)[l]{$+ \frac{1}{2}|\Phi^{+}\rangle (\alpha|1\rangle + \beta |0\rangle)$ }}
\put(60,5.){\makebox(0,0)[l]{$+ \frac{1}{2}|\Phi^{-}\rangle (\alpha|1\rangle - \beta |0\rangle)$ }}
\normalsize

\put(0,-12){\makebox(0,0)[l]{(c)}}

\multiput(37,48)(0,-4){33}{\line(0,-1){2}}
\multiput(37.2,48)(0,-4){33}{\line(0,-1){2}}

\end{picture}
\end{figure}

\vspace*{-0.5cm}

\begin{eqnarray*} 
	\mbox{Alice finds} |\Psi^{+}\rangle &\longrightarrow^0& \mbox{Bob does nothing}\\
	\mbox{Alice finds} |\Psi^{-}\rangle &\longrightarrow^1& \mbox{Bob performs}\; \sigma_z\\
	\mbox{Alice finds} |\Phi^{+}\rangle &\longrightarrow^2& \mbox{Bob performs}\; \sigma_x\\
	\mbox{Alice finds} |\Phi^{-}\rangle &\longrightarrow^3& \mbox{Bob performs}\; \sigma_z\sigma_x\\
\end{eqnarray*}

\begin{figure}[hbt]

\setlength{\unitlength}{0.9mm}
\begin{picture}(75,3)

\thicklines

\put(0,2){\makebox(0,0)[l]{(d)}}

\put(13,2){\circle{5}}
\put(25.2,2){\circle{5}}
\put(49.2,2){\circle{5}}
\put(47.7,2){\makebox(0,0)[l]{$\small\psi$}}
\put(23.7,3){\line(1,0){3}}
\put(23.7,1){\line(1,0){3}}
\put(11.5,3){\line(1,0){3}}
\put(11.5,1){\line(1,0){3}}

\multiput(16.9,2.1)(2.33,0.){3}{\makebox(0,0)[c]{$\bf\sim$}}
\multiput(16.85,1.9)(2.33,0.){3}{\makebox(0,0)[c]{$\bf\sim$}}

\end{picture}
\vspace*{1.25cm}
\caption{The basic steps of quantum state teleportation. Alice and Bob are spatially
separated, Alice on the left of the dashed line, Bob on the right. (a) Alice and Bob 
share a maximally entangled pair of particles in the state 
$(|00\rangle + |11\rangle)/\protect\sqrt{2}$. Alice wants to teleport the unknown 
state $|\psi\rangle$ to Bob. (b) The total state of the three particles that Alice and Bob
are holding is rewritten in the Bell basis Eqs. (\protect\ref{Bell1}-\ref{Bell}) 
for the two particles Alice is holding. Alice performs a measurement 
that projects the state of her two particles onto one of the four Bell states. 
(c) She transmits the result encoded in the numbers $0,1,2,3$ to Bob, who performs 
a unitary transformation ${\bf 1},\sigma_z,\sigma_x,\sigma_z\sigma_x$ that depends 
only on the measurement result that Alice obtained but {\bf not} on the state $|\psi\rangle$! 
(d) After Bob has applied the appropriate unitary operation on his particle
he can be sure that he is now holding the state that Alice was holding in (a).}
\label{teleport}
\end{figure}

\newpage

\newpage

\newpage

\newpage

\newpage

\newpage

\newpage

\begin{thebibliography}{99}
%
\bibitem{Aspect82} A. Aspect, J. Dalibard, P. Grangier, and 
G. Roger, Phys. Rev. Lett.
{\bf 49}, 1804 (1982) 
%
\bibitem{Barenco96} A. Barenco, Cont. Phys. {\bf 37}, 375 (1996).
%
\bibitem{Bell65} J.S. Bell, Physics {\bf 1}, 195 (1965)
%
\bibitem{Bell66} J.S. Bell, Rev. Mod. Phys. {\bf 38}, 447 (1966)
%
\bibitem{Bell87} J. S. Bell, {\em Speakable and unspeakable in quantum 
mechanics} (Cambridge Unicersity Press, 1987).
%
\bibitem{Bennett93} C.H. Bennett, G. Brassard, C. Crepeau, R. Jozsa, A. Peres, 
and W.K. Wootters, Phys. Rev. Lett. {\bf 70}, 1895 (1993) 
%
\bibitem{Bennett96a} C.H. Bennett, H.J. Bernstein, S. Popescu, and B. Schumacher,
Phys. Rev. A {\bf 53}, 2046 (1996)
%
\bibitem{Bennett96b} C.H. Bennett, G. Brassard, S. Popescu, B. Schumacher, 
J.A. Smolin, and W.K. Wootters, Phys. Rev. Lett. {\bf 76}, 722 (1996)
%
\bibitem{Bennett96c} C.H. Bennett, D.P. DiVincenzo, J.A. Smolin, and 
W.K. Wootters, Phys. Rev. A {\bf 54}, 3824 (1996)
%
\bibitem{deMartini98} D. Boschi, S. Branca, F. DeMartini, L. Hardy, and S. Popescu, Phys. Rev. Lett. {\bf 80}, 1121 (1998)
%
\bibitem{Bose98} S. Bose, V. Vedral and P. L. Knight, Phys. Rev. A. 
{\bf 57}, 822 (1998).
%
%
\bibitem{Calderbank96} A.R. Calderbank and P.W. Shor, Phys. Rev. A {\bf 54}, 1098 (1996)
%
\bibitem{Clauser78} J.F. Clauser and A. Shimony, Rep. Prog. Phys. {\bf 41}, 1881 (1978)
%
\bibitem{Grover97} L.K. Grover, lanl e-print quant-ph/9704012.
%
\bibitem{Horodecki97} M. Horodecki, P. Horodecki, R. Horodecki, Phys. Rev. Lett. {\bf 78},
574 (1997)
%
\bibitem{Huelga98} J.I. Cirac, A. Ekert, S.F. Huelga, and C. Macchiavello, 
preprint
%
\bibitem{Cover91} T. M. Cover and J. A. Thomas, {\em Elements of Information
Theory} (John Wiley and Sons Inc., New York, 1991). 
%
\bibitem{Deutsch96} D. Deutsch, A. Ekert, R. Jozsa, C. Macchiavello, 
S. Popescu, A. Sanpera, Phys. Rev. Lett. {\bf 77}, 2818 (1996)
%
\bibitem{Ekert96} A. Ekert and R. Jozsa, Rev. Mod. Phys. {\bf 68}, 733 (1996).
%
\bibitem{Chiara96} A. Ekert and C. Macchiavello, Phys. Rev. Lett. {\bf 77}, 2585 (1996)
%
\bibitem{Gisin96} N. Gisin, Phys. Lett. {\bf 210}, 151 (1996)
%
\bibitem{Horodecki97} M. Horodecki, P. Horodecki, and R. Horodecki, Phys. Rev. Lett. {\bf 78}, 574 (1997)
%
\bibitem{Horodecki97a} M. Horodecki and R. Horodecki, lanl e-print quant-ph/9705003
%
\bibitem{Huelga97} S.F. Huelga, C. Macchiavello, T. Pellizzari, A.K. Ekert, M.B. Plenio, and J.I. Cirac, Phys. Rev. Lett. {\bf 79}, 3865 (1997)
%
\bibitem{Jozsa97} R. Jozsa, eprint quant-ph/9707034
%
\bibitem{Lo97} H.-W. Lo and S. Popescu, lanl e-print quant-ph/9707038
%
\bibitem{Popescu98} N. Linden, S. Massar, and S. Popescu, preprint
%
\bibitem{Murao98} M. Murao, M.B. Plenio, S. Popescu, V. Vedral, and P.L. Knight,
to appear in Phys. Rev. A. 
%
\bibitem{Plenio96} M.B. Plenio and P.L. Knight, Phys. Rev. A {\bf 53}, 2986 (1996)
%
\bibitem{Plenio97} M.B. Plenio and P.L. Knight, Proc. R. Soc. Lond. A {\bf 453}, 2017 (1997)
%
\bibitem{Popescu94} S. Popescu, Phys. Rev. Lett. {\bf 72}, 797 (1994)
%
\bibitem{Popescu97} S. Popescu and D. Rohrlich, Phys. Rev. A {\bf 56}, R3319 (1997)
%
\bibitem{Santos} E. Santos, Phys. Rev. Lett. {\bf 66}, 1388 (1991)
%
\bibitem{Schumacher95} B. Schumacher, Phys. Rev. A {\bf 51}, 2738 (1995)
%
\bibitem{Shor95} P.W. Shor, Phys. Rev. A {\bf 52}, 2493 (1995)
%
\bibitem{Steane96} A.M. Steane, Proc. R. Soc. Lond. A {\bf 452}, 2551 (1996)
%
\bibitem{Vedral97a} V. Vedral, M.B. Plenio, and M.A. Rippin, and P.L. Knight,
Phys. Rev. Lett. {\bf 78}, 2275 (1997)
%
\bibitem{Vedral97b} V. Vedral, M.A. Rippin and M. B. Plenio, J. Mod. Opt. {\bf 44}, 2185 (1997)
%
\bibitem{Vedral97c} V. Vedral, M.B. Plenio, K. Jacobs, and P.L. Knight,
Phys. Rev. A {\bf 56}, 4452 (1997)
%
\bibitem{VP98a} V. Vedral and M. B. Plenio, Phys. Rev. A {\bf 57}, 1619 (1998).
%
\bibitem{VP98b} V. Vedral and M. B. Plenio, invited review to appear in 
Prog. Quant. Electr. (1998).
%
\bibitem{Wootters82} W.K. Wootters and W.H. Zurek, Nature {\bf 299}, 802 (1982).
%
\bibitem{Wootters98} W.K. Wootters, Phys. Rev. Lett. {\bf 80}, 2245 (1998).
%
\bibitem{Zeilinger98} D. Bouwmeester, J.W. Pan, K. Mattle, M. Eibl, H. Weinfurter, and A. Zeilinger, Nature {\bf 390}, 575 (1997)
%
\bibitem{Einstein} A. Einstein, this quote is attributed to Einstein, however,
we were unable to trace the original reference.
%
\end{thebibliography}
\end{document}